\documentclass[aps,10pt,prb,twocolumn,superscriptaddress]{revtex4-2}
\usepackage{graphicx}
\usepackage{amsmath}
\usepackage{amsfonts}
\usepackage{amssymb,mathrsfs}
\usepackage{upgreek}
\usepackage{bbold,yfonts}

\newcommand{\bs}[1]{\boldsymbol{#1}}

\begin{document}

\title{Hydrodynamic collective modes in graphene}

\author{B.N. Narozhny}
\affiliation{\mbox{Institut for Theoretical Condensed Matter Physics, Karlsruhe Institute of 
Technology, 76128 Karlsruhe, Germany}}
\affiliation{National Research Nuclear University MEPhI (Moscow Engineering Physics Institute),
  115409 Moscow, Russia}
\author{I.V. Gornyi}
\affiliation{\mbox{Institut for Theoretical Condensed Matter Physics, Karlsruhe Institute of
Technology, 76128 Karlsruhe, Germany}}
\affiliation{\mbox{Institut for Quantum Materials and Technologies, Karlsruhe Institute of
Technology, 76021 Karlsruhe, Germany}}
\affiliation{Ioffe Institute, 194021 St. Petersburg, Russia}
\author{M. Titov}
\affiliation{Radboud University Nijmegen, Institute for Molecules and Materials, NL-6525 AJ
Nijmegen, The Netherlands}

\date{\today}

\begin{abstract}
  Collective behavior is one of the most intriguing aspects of the
  hydrodynamic approach to electronic transport. Here we provide a
  consistent, unified calculation of the dispersion relations of the
  hydrodynamic collective modes in graphene. Taking into account
  viscous effects, we show that the hydrodynamic sound mode in
  graphene becomes overdamped at sufficiently large momentum
  scales. Extending the linearized theory beyond the hydrodynamic
  regime, we connect the diffusive hydrodynamic charge density
  fluctuations with plasmons. 
\end{abstract}

\maketitle

Electronic hydrodynamics is quickly growing into a mature field of
solid state physics
\cite{rev,luc,geim1,kim1,mac,geim2,kim2,geim3,geim4,ihn,goo,gal,imh,imm,hydro1,me1,msf}.
Similarly to the usual hydrodynamics \cite{dau6}, this approach offers
a universal, long-wavelength description of collective flows in
interacting many-electron systems. Such flows have been experimentally
confirmed \cite{geim2} to be more efficient than the usual
single-electron (ballistic or diffusive) transport.

In graphene, hydrodynamic collective modes have been
considered by many authors
\cite{luc,schutt,cosmic,lev13,hydro1,fog,svin,ldsp,ks20,fat}. All of them agree
that at charge neutrality, the ideal electronic fluid (i.e., neglecting
all dissipative processes) allows for a sound-like collective mode
(which has been referred to as either the ``cosmic sound''
\cite{cosmic} or the ``second sound'' \cite{ks20}) with the dispersion
relation
\begin{equation}
\label{cs0}
\omega = v_gq/\sqrt{2},
\end{equation}
where $v_g$ is the quasiparticle velocity in graphene. Taking into
account dissipation changes the above dispersion relation giving rise
to damping. To the best of our knowledge, no consensus on the latter
effect has been reached so far with several contradicting results
available in the literature \cite{hydro1,svin}.

The hydrodynamic approach to electronic systems is applicable in an
intermediate parameter regime \cite{rev,luc}. In particular, the
underlying gradient expansion is valid at length scales much larger
than the typical length scale $\ell_{ee}$ describing the energy- and
momentum-conserving interaction (responsible for equilibration of the
system). At smaller length scales, one can study more traditional
collective excitations in interacting many-electron systems, including
plasmons
\cite{hydro1,Giuliani,ldsp,ks20,fat,hill09,prin11,fei12,chen12,lev13,bas,kop17,kop18,polkop20,pol20pl,kop20p1,kop20p2,mach19,per20,mach20,falko20},
which behavior is well established both theoretically and
experimentally.

\begin{figure}[t]
\centerline{\includegraphics[width=0.4\textwidth]{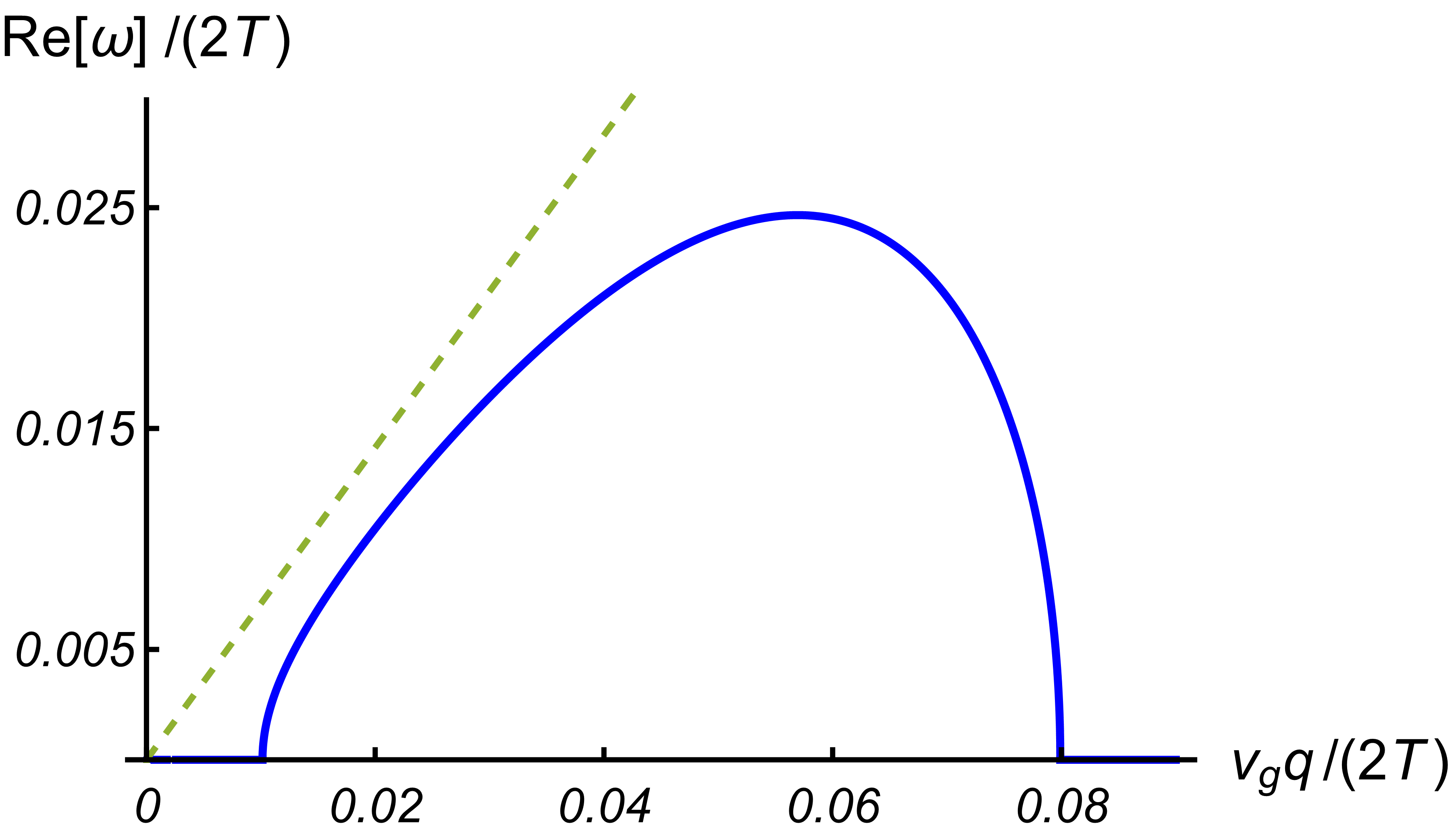}}
\bigskip
\centerline{\includegraphics[width=0.4\textwidth]{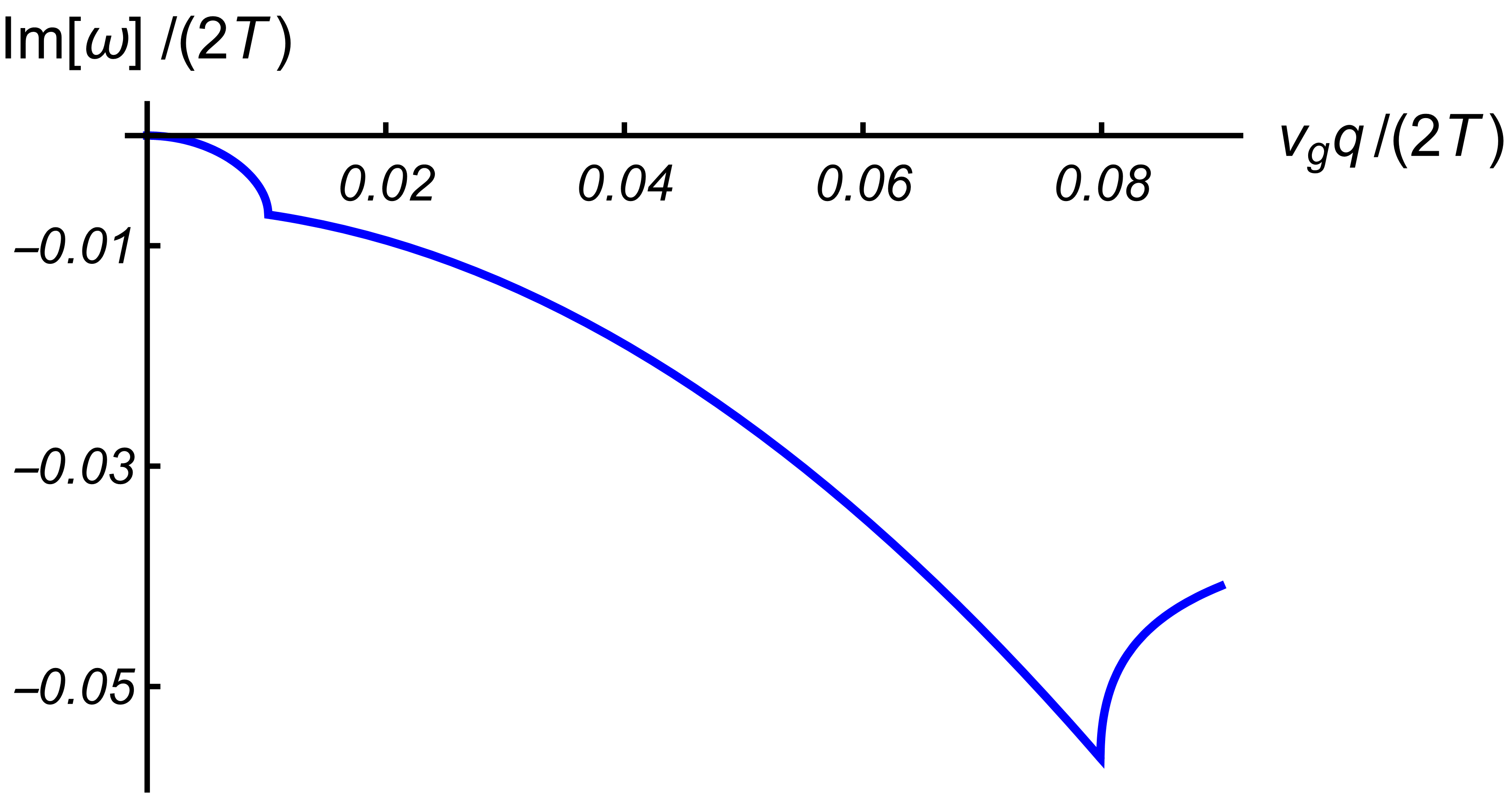}
}
\caption{Real (top) and imaginary (bottom) parts of the hydrodynamic
  sound dispersion in neutral graphene taking into account viscosity
  and weak disorder, Eq.~(\ref{csv}). The numerical values were
  computed with the realistic parameter values taken from
  Refs.~\onlinecite{geim1,geim4,gal}; see the main text. The
  dispersion acquires a finite real part at the threshold value of
  momentum determined by dissipation. The mode becomes overdamped at
  small enough momenta, still in the region of the growing real
  part. The dashed line shows the ideal dispersion, Eq.~(\ref{cs0}). }
\label{fig1:sd0v}
\end{figure}

In this paper we provide a consistent, unified calculation of the
dispersion relations of the hydrodynamic collective modes in
graphene. While the true hydrodynamics is universal (as long as no
symmetries are broken), graphene is somewhat unique in the sense that
there are two length scales associated with electron-electron
interaction that are parametrically different in the weak coupling
limit \cite{hydro0,luc,mfss,ks20}. This allows us to extend the
results of the linearized hydrodynamic theory \cite{hydro0,hydro1,me3}
to the length scales smaller that $\ell_{ee}$ (going beyond the
small-momentum expansion of Ref.~\onlinecite{hydro1}). At that point
the sound mode (\ref{cs0}) in neutral graphene (see
Fig.~\ref{fig1:sd0v}) becomes overdamped due to the high viscosity
\cite{geim1,geim4,me2}
\begin{equation}
\label{csv}
\omega = 
\sqrt{\frac{v_g^2q^2}{2} - \frac{\left(1\!+\!q^2\ell_G^2\right)^2}{4\tau^2_{{\rm dis}}}}
- i \frac{1\!+\!q^2\ell_G^2}{2\tau_{{\rm dis}}},
\end{equation}
where $\tau_{\rm dis}$ is the disorder mean free time and $\ell_G$ is
the so-called Gurzhi length \cite{moo,pol17,mr2,cfl,sven1} (here $\nu$
stands for the kinematic viscosity \cite{luc,geim1,geim4,me2})
\begin{equation}
\label{gl}
\ell_G = \sqrt{\nu\tau_{\rm dis}}.
\end{equation}
This mode describes energy fluctuations and is completely decoupled
from charge fluctuations. The latter are purely diffusive within the
hydrodynamic approach, where dissipation is described by the momentum-
and frequency-independent coefficients, including the electrical
conductivity and viscosity. 

Extending the linearized theory beyond the hydrodynamic regime, we are
able to connect the charge fluctuations with the more conventional
plasmons by taking into account the frequency and momentum dependence
of conductivity. At charge neutrality we find the plasmon mode
\begin{equation}
\label{pd0}
\omega = 
\sqrt{\frac{v_g^2\varkappa q}{2}\left(1 \!+\! \frac{q}{\varkappa}\right)
- \frac{v_g^4\varkappa^2}{64\pi^2\sigma_0^2} }-i\frac{v_g^2\varkappa}{8\pi\sigma_0},
\end{equation}
where $\sigma_0$ is the conductivity in neutral graphene
\cite{rev,luc,me1,kash}
\begin{equation}
\label{sq}
\sigma_0 = \frac{2e^2T\ln2}{\pi} \frac{\tau_{11}\tau_{\rm dis}}{\tau_{11}\!+\!\tau_{\rm dis}},
\end{equation}
and $\varkappa$ is the inverse Thomas-Fermi screening
length. Neglecting dissipation and for small momenta, the dispersion
(\ref{pd0}) coincides with the result of Ref.~\onlinecite{schutt}.

Finally, we extend our results over the whole range
of carrier densities up to the degenerate (``Fermi-liquid'')
regime. Given the weak density dependence of the kinematic viscosity
in graphene \cite{geim1,geim4,me2} the sound dispersion remains
qualitatively similar to that shown in Fig.~\ref{fig1:sd0v} at all
doping levels.

\section{Hydrodynamic theory of electronic transport in graphene}
\label{ht}

In this Section, we briefly review the hydrodynamic theory of
electronic transport in graphene.

\subsection{Nonlinear hydrodynamic equations}

The complete set of hydrodynamic
equations includes the generalized Navier-Stokes equation \cite{me1,msf}
\begin{subequations}
\label{hydro}
\begin{eqnarray}
\label{eq1g}
&&
\!\!\!\!\!\!
W(\partial_t+\bs{u}\!\cdot\!\bs{\nabla})\bs{u}
+
v_g^2 \bs{\nabla} P
+
\bs{u} \partial_t P 
+
e(\bs{E}\!\cdot\!\bs{j})\bs{u} 
=
\\
&&
\nonumber\\
&&
=
v_g^2 
\left[
\eta \Delta\bs{u}
-
\eta_H \Delta\bs{u}\!\times\!\bs{e}_B
+
en\bs{E}
+
\frac{e}{c} \bs{j}\!\times\!\bs{B}
\right]
-
\frac{W\bs{u}}{\tau_{{\rm dis}}},
\nonumber
\end{eqnarray}
the continuity equations \cite{rev,luc,me1}
\begin{equation}
\label{cen1}
\partial_t n + \bs{\nabla}\!\cdot\!\bs{j} = 0,
\end{equation}
\begin{equation}
\label{ceni1}
\partial_t n_I + \bs{\nabla}\!\cdot\!\bs{j}_I = - \frac{n_I\!-\!n_{I,0}}{\tau_R},
\end{equation}
and the generalized ``heat transport'' equation \cite{alf,lev19,meig1}
(we follow the usual approach \cite{dau6} using the entropy flow
equation instead of the continuity equation for energy).
\begin{eqnarray}
\label{eqent}
&&
T\left[\frac{\partial s}{\partial t}
+
\bs{\nabla}\!\cdot\!
\frac{3P\bs{u} - \mu\bs{j} - \mu_I\bs{j}_I}{T}\right]
=
\\
&&
\nonumber\\
&&
\qquad\qquad
=
\delta\bs{j}\!\cdot\!
\left[e\bs{E}\!+\!\frac{e}{c}\bs{u}\!\times\!\bs{B}\!-\!T\bs{\nabla}\frac{\mu}{T}\right]
-
T\delta\bs{j}_I\!\cdot\!\bs{\nabla}\frac{\mu_I}{T}
\nonumber\\
&&
\nonumber\\
&&
\qquad\qquad\qquad
+
\frac{\eta}{2}\left(\nabla_\alpha u_\beta \!+\! \nabla_\beta u_\alpha
\!-\! \delta_{\alpha\beta} \bs{\nabla}\!\cdot\!\bs{u}\right)^2
\nonumber\\
&&
\nonumber\\
&&
\qquad\qquad\qquad
-
\frac{n_E\!-\!n_{E,0}}{\tau_{RE}}
+
\mu_I \frac{n_I\!-\!n_{I,0}}{\tau_R}
+
\frac{W\bs{u}^2}{v_g^2\tau_{\rm dis}}.
\nonumber
\end{eqnarray}
\end{subequations}
Here $\bs{u}$ is the hydrodynamic velocity, $c$ is the speed of light,
and $n$ and $n_I$ are the carrier and imbalance densities ($n_{I,0}$
is the equilibrium value), related to the quasiparticle densities in
each of the two bands by
\[
n = n_+ - n_-, \qquad
n_I = n_+ + n_-.
\]
The carrier density $n$ differs from the charge density by a
multiplicative factor of the electric charge, $e$. Similarly, we define
the two quasiparticle currents, $\bs{j}$ and $\bs{j}_I$,
\[
\bs{j} = \bs{j}_+ - \bs{j}_-,
\qquad
\bs{j}_I = \bs{j}_+ + \bs{j}_-,
\]
with the electric current $\bs{J}=e\bs{j}$. We also define the
two chemical potentials, $\mu$ and $\mu_I$,
\[
\mu = (\mu_++\mu_-)/2,
\qquad
\mu_I = (\mu_+-\mu_-)/2,
\]
allowing for the two independent chemical potentials for each band out
of equilibrium \cite{alf} (hence the term ``imbalance''). The
remaining vector quantities in Eqs.~(\ref{hydro}) are the electric
field $\bs{E}$ and the magnetic field $\bs{B}$. The thermodynamic
quantities are the enthalpy density $W$, pressure $P$, entropy density
$s$, and temperature $T$. Finally, $\eta$ and $\eta_H$ are the shear
and Hall viscosities, $\tau_R$ is the recombination time \cite{alf}
[the recombination term in Eq.~(\ref{ceni1}) agrees with
  Ref.~\onlinecite{lev19}, whereas Ref.~\onlinecite{alf} suggests a
  slightly different term that is proportional to $\mu_I$ instead of
  the $\delta n_I$], and $\tau_{RE}$ is the energy relaxation time
\cite{meig1}. In equilibrium, $\mu_I=0$.

In comparison to the usual hydrodynamics \cite{dau6}, the electronic
system in graphene is characterized by one additional variable
describing the second band. Traditional ideal fluid is described by
two thermodynamic variables, e.g., density and pressure, and the
velocity field. As a result, in two dimensions one needs four
equations to describe the dynamics of the flow. Two of these are given
by the Euler equation, the third is the continuity equation, while the
fourth can be either the continuity equation for energy or the
adiabaticity equation (i.e., the continuity equation for entropy). In
graphene these are Eqs.~(\ref{eq1g}), (\ref{cen1}), and (\ref{eqent})
in the absence of dissipation. The additional continuity equation
(\ref{ceni1}) for the quasiparticle density $n_I$ appears exactly due
to the presence of the second band, which is why the overall number of
hydrodynamic equations as well as independent variables in graphene is
five. As the additional variable one can choose either $n_I$ or the
corresponding chemical potential $\mu_I$.

The entropy flow equation (\ref{eqent}) should be compared to the
corresponding equations in Refs.~\onlinecite{luc,alf,lev19}. The four
equations contain mostly the same terms (up to trivial notation
changes) with the following exceptions. Equation (54) of
Ref.~\onlinecite{luc} is written in the relativistic notation omitting
the imbalance mode, quasiparticle recombination, and disorder
scattering, all of which are discussed separately elsewhere in
Ref.~\onlinecite{luc}. Reference~\onlinecite{alf} was the first to
focus on the imbalance mode with Eq.~(2.6) containing all the terms of
Eq.~(\ref{eqent}) except for the viscous term. Finally, Eq.~(1c) of
Ref.~\onlinecite{lev19} contains all of the terms in Eq.~(\ref{eqent})
and in addition contains a term describing energy relaxation due to
electron-phonon scattering that is neglected in this paper
(generalization of the resulting theory is straightforward).

Weak disorder scattering is described in Eqs.~(\ref{eq1g}) and
(\ref{eqent}) by the mean free time $\tau_{\rm dis}$. The disorder
contribution to the hydrodynamic equations was derived in
Ref.~\onlinecite{me1} using the simplest $\tau$-approximation to the
kinetic equation. A better version of the disorder collision integral
in graphene should involve the Dirac factors suppressing
backscattering \cite{kash18} which would lead to the similar
approximation but with the transport scattering time. In graphene,
this brings about a factor of $2$. In this paper, we treat $\tau_{\rm
  dis}$ as a phenomenological parameter adopting the approach of
Ref.~\onlinecite{gal}.

The imbalance density $n_I$ appears under the assumption of the
approximate conservation of the number of particles in each individual
band. The processes that break this conservation (i.e., mix electrons
and holes) involve the three-particle scattering, Auger processes
\cite{alf}, and most importantly, impurity assisted electron-phonon
coupling \cite{srl}. These effects are described in Eq.~(\ref{ceni1})
by the phenomenological \cite{meg,mrexp} recombination time
\cite{mr1}, $\tau_R$, as well as the energy relaxation time
$\tau_{RE}$ in Eq.~(\ref{eqent}).

\subsection{Dissipative corrections to quasiparticle currents}

The usual hydrodynamic flow \cite{dau6} is a mass flow where
dissipative processes lead to a correction to the energy flux as
described by the thermal conductivity. Consequently the flow is
characterized by three dissipative coefficients, the thermal
conductivity $\varkappa$ and two viscosities $\eta$ and $\zeta$. In
contrast, electronic hydrodynamics in graphene describes an energy
flow where the quasiparticle currents acquire dissipative
corrections. The energy flow is proportional to the momentum density
and hence can only be affected by disorder, which is ``extrinsic'' to
the hydrodynamic theory. As a result, the dissipative coefficients
include the electrical conductivity $\sigma$ and viscosity, while the
thermal conductivity has to be computed by solving the linear response
equations (similarly to the electrical conductivity in the standard
theory). Within the three-mode approximation of Ref.~\onlinecite{me1},
the bulk viscosity vanishes, ${\zeta=0}$. In the absence of the
magnetic field the dissipative corrections are related to external
bias by means of a ``conductivity matrix'' \cite{me1,alf,lev19}
\begin{equation}
\label{djs}
\begin{pmatrix}
\delta\bs{j} \cr
\delta\bs{j}_I
\end{pmatrix}
=\widehat\Sigma
\begin{pmatrix}
e\bs{E} - T\bs{\nabla}(\mu/T) \cr
-T\bs{\nabla}(\mu_I/T)
\end{pmatrix}.
\end{equation}
In particular, at the Dirac point ${\mu=\mu_I=0}$ the matrix
$\widehat\Sigma$ is diagonal with the upper diagonal element defining
(in the absence of disorder) the ``quantum'' or ``intrinsic''
conductivity \cite{me1,luc,rev,alf,lev19}
\begin{equation}
\sigma_Q = e^2 \Sigma_{11}(0).
\end{equation}
In the hydrodynamic theory of graphene, the elements of the matrix
$\widehat\Sigma$ play the role that is equivalent to that of the
thermal conductivity $\varkappa$ in the usual hydrodynamics. The
matrix nature of $\widehat\Sigma$ reflects the band structure of
graphene. In the case of strong recombination, the imbalance mode
becomes irrelevant and one is left with the single dissipative
coefficient $\sigma_Q$, see Ref.~\onlinecite{luc}.

\begin{figure}[t]
\centerline{\includegraphics[width=0.45\textwidth]{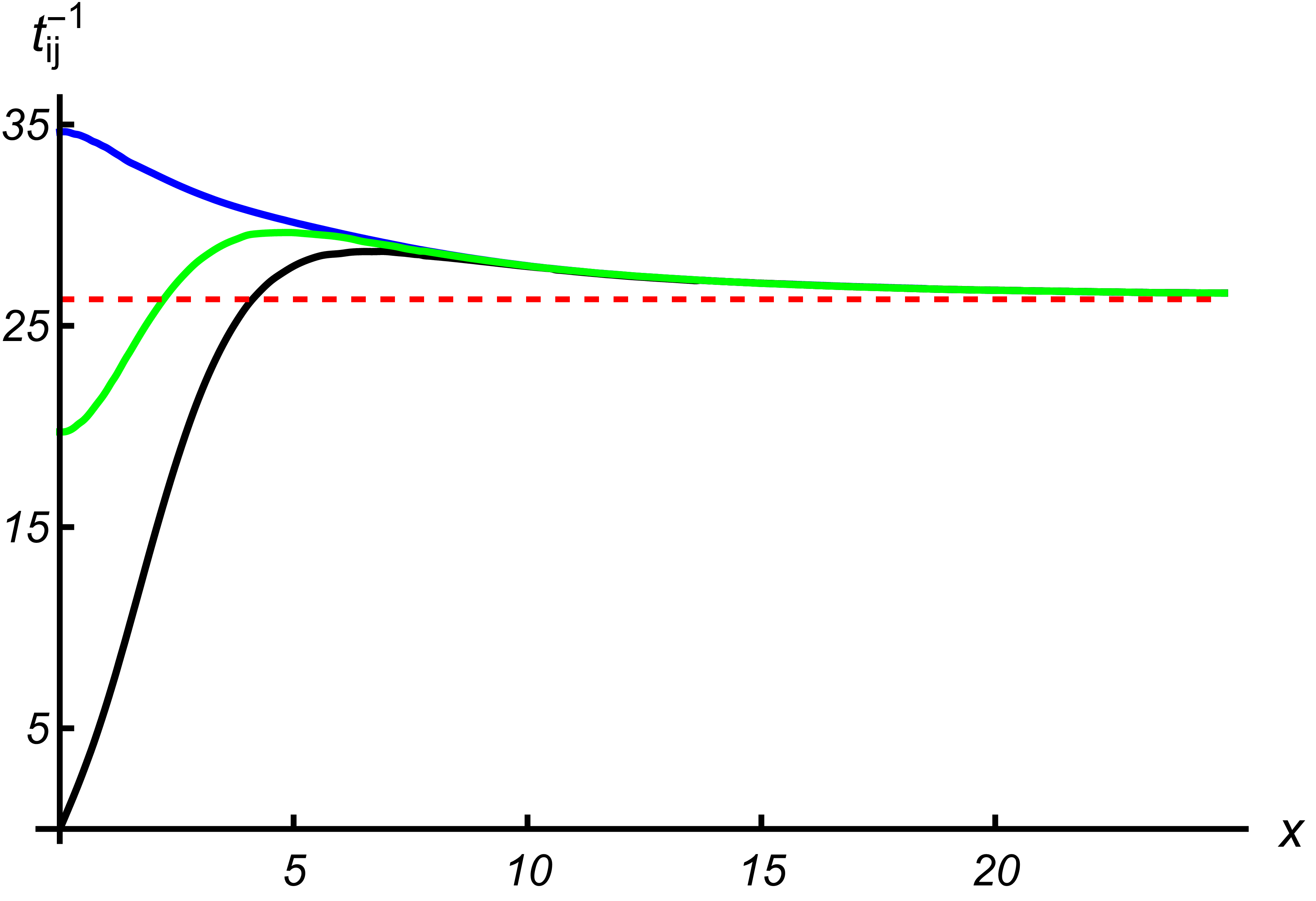}
}
\caption{Dimensionless scattering rates comprising the matrix
  $\widehat{\textswab{T}}$: $t_{11}^{-1}$, $t_{12}^{-1}$,
  $t_{22}^{-1}$ (blue, black, and green, respectively). The red dashed
  line indicates the ``Fermi-liquid'' limit, Eq.~(\ref{tfl}).}
\label{fig2:tij}
\end{figure}

\subsubsection{Macroscopic currents within the three-mode approximation}

Within the three-mode approximation of Ref.~\onlinecite{me1}, one
defines three macroscopic currents (using $\bar{W}=3\bar{n}_E/2$)
\begin{equation}
\label{jlr}
\bs{j} = \bar{n}\bs{u} \!+\! \delta\bs{j},
\quad
\bs{j}_I = \bar{n}_I\bs{u} \!+\! \delta\bs{j}_I,
\quad
\bs{j}_E = \frac{3}{2}\bar{n}_E\bs{u},
\end{equation}
where $\bar{n}$, $\bar{n}_I$, and $\bar{n}_E$ are the equilibrium
values of the carrier, imbalance, and energy densities, respectively.
The linear response theory relates the dissipative corrections
$\delta\bs{j}$ and $\delta\bs{j}_I$ to the external bias by
Eq.~(\ref{djs}). The dimensionless conductivity matrix (at
${\bs{B}=0}$) is given by \cite{me1}
\begin{subequations}
\label{sigma3}
\begin{equation}
\label{sigmam}
\widehat\Sigma=
\widehat{\textswab{M}}\,
\widehat{\textswab{S}}_{xx}^{-1}
\widehat{\textswab{M}},
\quad
\widehat{\textswab{S}}_{xx} = \frac{\alpha_g^2T^2}{2{\cal T}^2}
\widehat{\textswab{T}}
+
\frac{\pi}{{\cal T}\tau_{\rm dis}}
\widehat{\textswab{M}},
\end{equation}
where
\begin{equation}
\label{mh}
\widehat{\textswab{M}}\!=\!
\begin{pmatrix}
1\!-\!\frac{2\tilde{n}^2}{3\tilde{n}_E}\frac{T}{\cal T} &
\frac{xT}{\cal T} \!-\! \frac{2\tilde{n}\tilde{n}_I}{3\tilde{n}_E}\!\frac{T}{\cal T}\cr
\frac{xT}{\cal T} \!-\!
\frac{2\tilde{n}\tilde{n}_I}{3\tilde{n}_E}\frac{T}{\cal T} & 
1\!-\!\frac{2\tilde{n}_I^2}{3\tilde{n}_E}\frac{T}{\cal T}
\end{pmatrix}\!,
\end{equation}
with dimensionless densities [see Eq.~(\ref{equilt}) below]
\begin{eqnarray}
\label{tiln}
&&
\tilde{n} = {\rm Li}_2\left(-e^{-x}\right) - {\rm Li}_2\left(-e^x\right)\!,
\quad
\tilde{n}_I=x^2/2+\pi^2/6,
\nonumber\\
&&
\nonumber\\
&&
\tilde{n}_E = -{\rm Li}_3\left(-e^x\right) 
- {\rm Li}_3\left(-e^{-x}\right)\!,
\\
&&
\nonumber\\
&&
x=\mu/T,
\quad
{\cal T} = 2T\ln\left[2\cosh(x/2)\right],
\nonumber
\end{eqnarray}
and dimensionless scattering rates
\begin{equation}
\label{taum}
\widehat{\textswab{T}}
=
\begin{pmatrix}
  t_{11}^{-1} & t_{12}^{-1}  \cr
  t_{12}^{-1} & t_{22}^{-1}
\end{pmatrix},
\qquad
t_{ij}^{-1}=\frac{8\pi{\cal T}}{\alpha_g^2NT^2} 
 \tau_{ij}^{-1},
\end{equation}
\end{subequations}
where $\tau_{ij}^{-1}$ are the scattering rates that can be obtained
by solving the kinetic equation within the three-mode approximation
\cite{hydro1,me1,mfss,ks20}. The zeros in the matrix (\ref{taum}) are
the manifestation of energy and momentum conservation, which is also
responsible for the vanishing dissipative correction to the energy
current in the absence of the magnetic field \cite{me1}. The three
dimensionless elements of the matrix $\widehat{\textswab{T}}$ are
shown in Fig.~\ref{fig2:tij} as a function of $x=\mu/T$.

\begin{figure}[t]
\centerline{\includegraphics[width=0.45\textwidth]{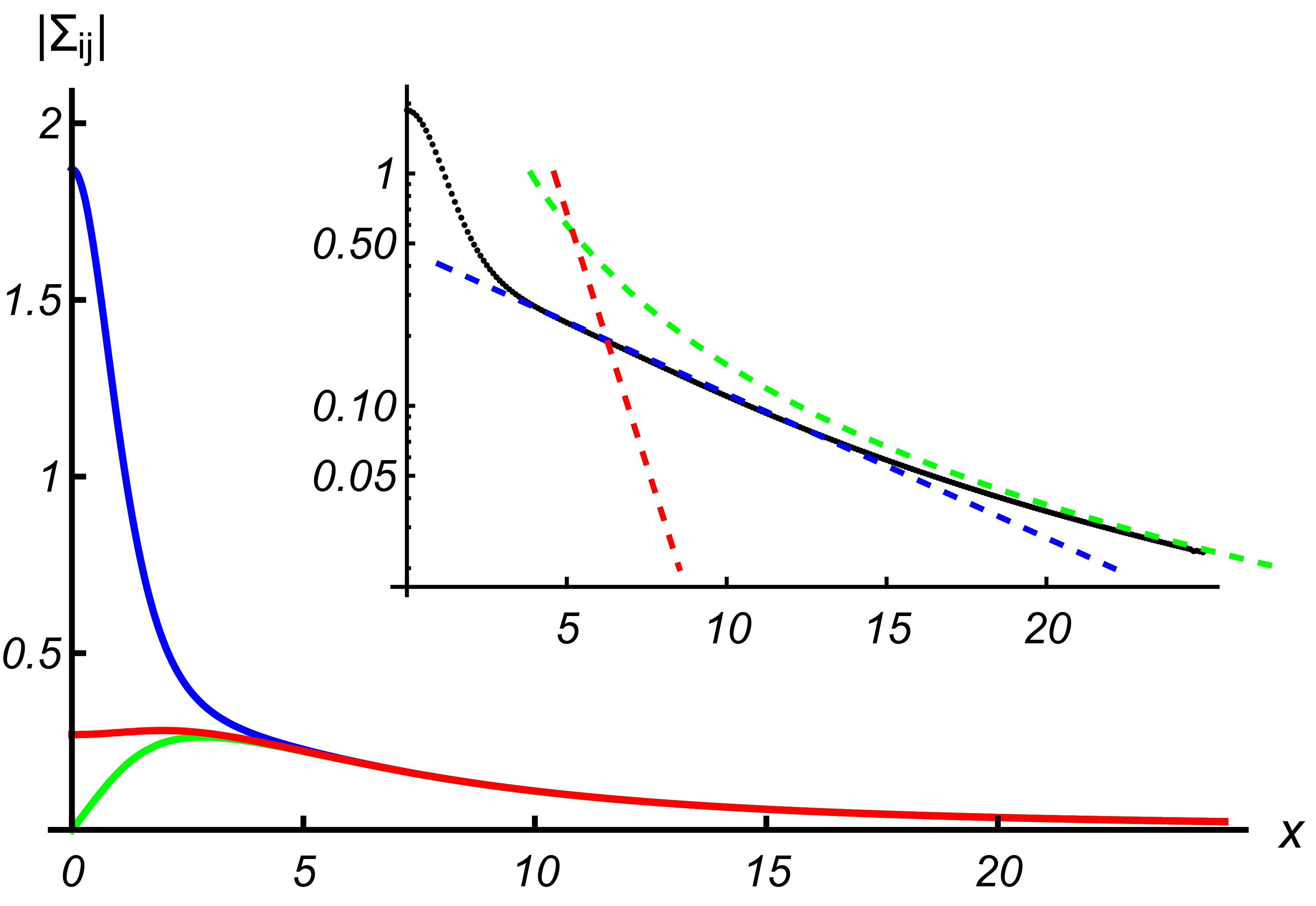}
}
\caption{Matrix elements of $\widehat\Sigma$. The blue, red, and green
  curves correspond to $\Sigma_{11}$, $|\Sigma_{12}|$,
  $|\Sigma_{22}|$, respectively (notice, that
  ${\Sigma_{21}=-\Sigma_{12}}$). The inset shows the log plot of
  $\Sigma_{11}$, where the red and blue lines indicate the exponential
  decay, while the green line is the power law $\sim x^{-2}$.}
\label{fig3:sij}
\end{figure}

The resulting matrix elements of the conductivity $\widehat\Sigma$ are
shown in Fig.~\ref{fig3:sij} as functions of $x=\mu/T$. As discussed
below, the numerical precision of the present calculation is
insufficient to track the exponential corrections to the scattering
rates in the degenerate regime. Hence, the decay shown in the inset in
Fig.~\ref{fig3:sij} might be an artifact.

\subsubsection{Dimensionless scattering rates}

In the degenerate regime all scattering rates (i.e., the matrix
elements $t_{ij}^{-1}$) coincide (up to exponentially small
corrections) approaching the limiting value
\begin{equation}
\label{tfl}
t_{ij}^{-1}(\mu\gg T) \rightarrow 8\pi^2/3.
\end{equation}

\begin{figure}[t]
\centerline{\includegraphics[width=0.4\textwidth]{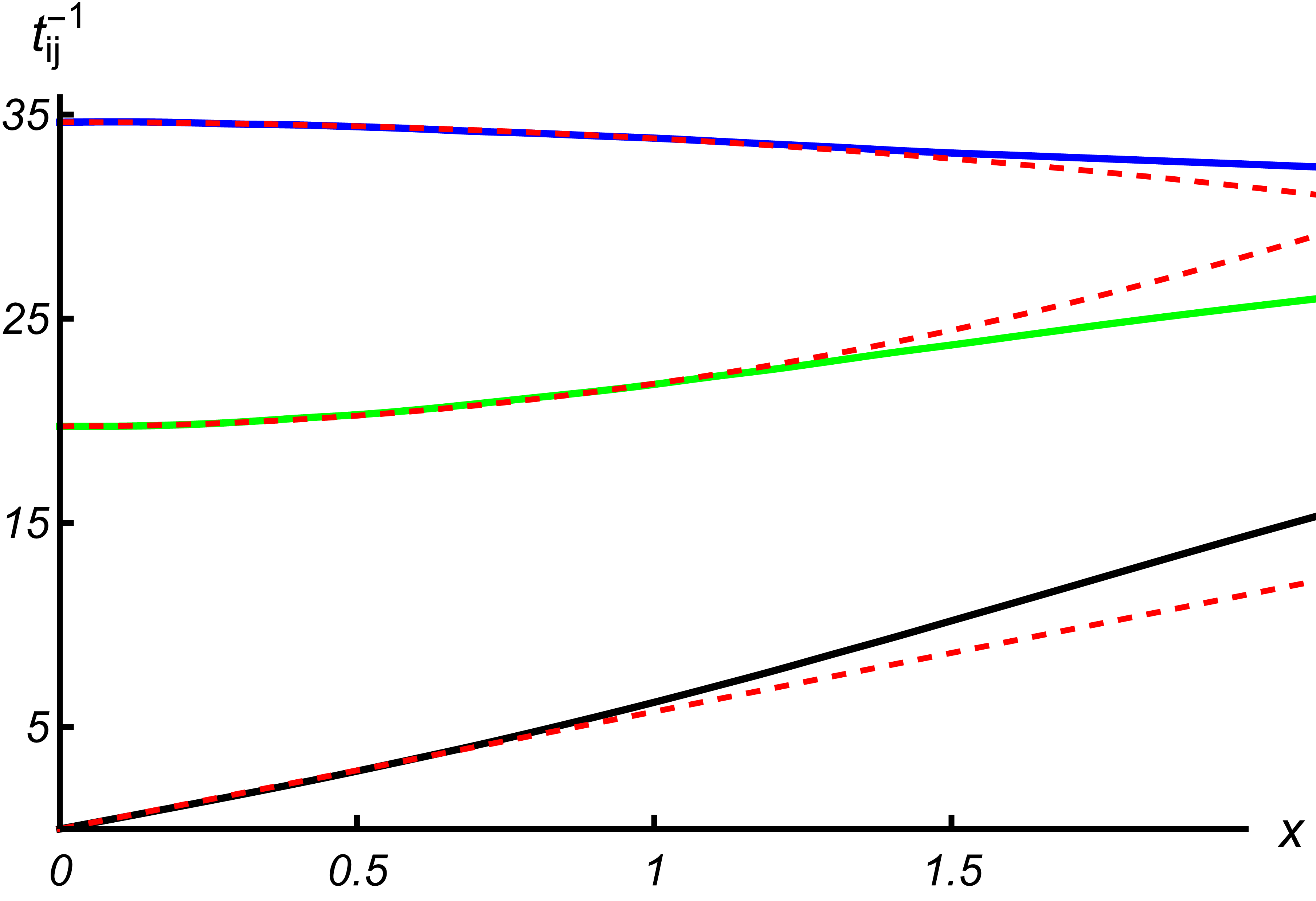}
}
\caption{Dimensionless scattering rates close to charge neutrality. The
  blue, black, and green curves correspond to $t_{11}^{-1}$,
  $t_{12}^{-1}$, $t_{22}^{-1}$, respectively. The red dashed lines
  indicate the leading behavior close to charge neutrality (\ref{tfl}).}
\label{fig4:tijdp}
\end{figure}

At $\mu=0$, the off-diagonal elements $t_{12}^{-1}(0)=0$, while the
diagonal elements $t_{ii}^{-1}(0)$ determine the diagonal elements of
the conductivity matrix, $\sigma_Q$ and $\sigma_I$, see below. For
small $x\ll1$ the dimensionless ``scattering rates'' $t_{ij}$ have the
form \cite{me3} (see Fig.~\ref{fig4:tijdp} for illustration)
\begin{subequations}
\label{tausdp}
\begin{equation}
\label{tau11dp}
\frac{1}{t_{11}} =\frac{1}{t_{11}^{(0)}} 
+ x^2\!\left(\frac{1}{t_{11}^{(2)}}\!-\!\frac{1}{8\ln2}\frac{1}{t_{11}^{(0)}}\right)\! 
+ {\cal O}(x^3),
\end{equation}
\begin{equation}
\label{tau12dp}
\frac{1}{t_{12}} = \frac{x}{t_{12}^{(1)}} + {\cal O}(x^3),
\end{equation}
\begin{equation}
\label{tau22dp}
\frac{1}{t_{22}} = \frac{1}{t_{22}^{(0)}} 
+ x^2\!\left(\frac{1}{t_{22}^{(2)}}\!-\!\frac{1}{8\ln2}\frac{1}{t_{22}^{(0)}}\right)\! 
+ {\cal O}(x^3).
\end{equation}
\end{subequations}
For unscreened Coulomb interaction, the dimensionless quantities
$t_{ij}^{(0,1,2)}$ are just numbers without any dependence on any
physical parameter. Numerically, one finds the following values
(neglecting the small \cite{kash} exchange contribution):
\[
\left(t_{11}^{(0)}\right)^{-1} \approx 34.63,
\quad
\left(t_{11}^{(2)}\right)^{-1} \approx 5.45,
\]
\[
\left(t_{12}^{(1)}\right)^{-1} \approx 5.72,
\quad
\left(t_{22}^{(0)}\right)^{-1} \approx 19.73,
\quad
\left(t_{22}^{(2)}\right)^{-1} \approx 5.65.
\]
Note that these values are slightly different from those listed in
Ref.~\onlinecite{me3}. The reason for this is the use of different
numerical methods. In the case of screened interaction, the quantities
$t_{ij}^{(0,1,2)}$ depend on the screening length.

\subsubsection{Conductivity matrix close to charge neutrality}

Close to charge neutrality we expand the matrix
$\widehat{\textswab{M}}$
\[
\widehat{\textswab{M}} = \widehat{\textswab{M}}(0) 
+ \delta\widehat{\textswab{M}} + {\cal O}(x^3), 
\]
with 
\begin{equation}
\label{m0}
\widehat{\textswab{M}}(0)
=
\begin{pmatrix}
1 & 0 \cr
0 & \delta_I 
\end{pmatrix}\!,
\end{equation}
where $\zeta(z)$ is the Riemann's zeta function and 
\[
\delta_I = 1-\frac{\pi^4}{162\zeta(3)\ln2}
\approx 0.28.
\]

\begin{figure}[t]
\centerline{\includegraphics[width=0.4\textwidth]{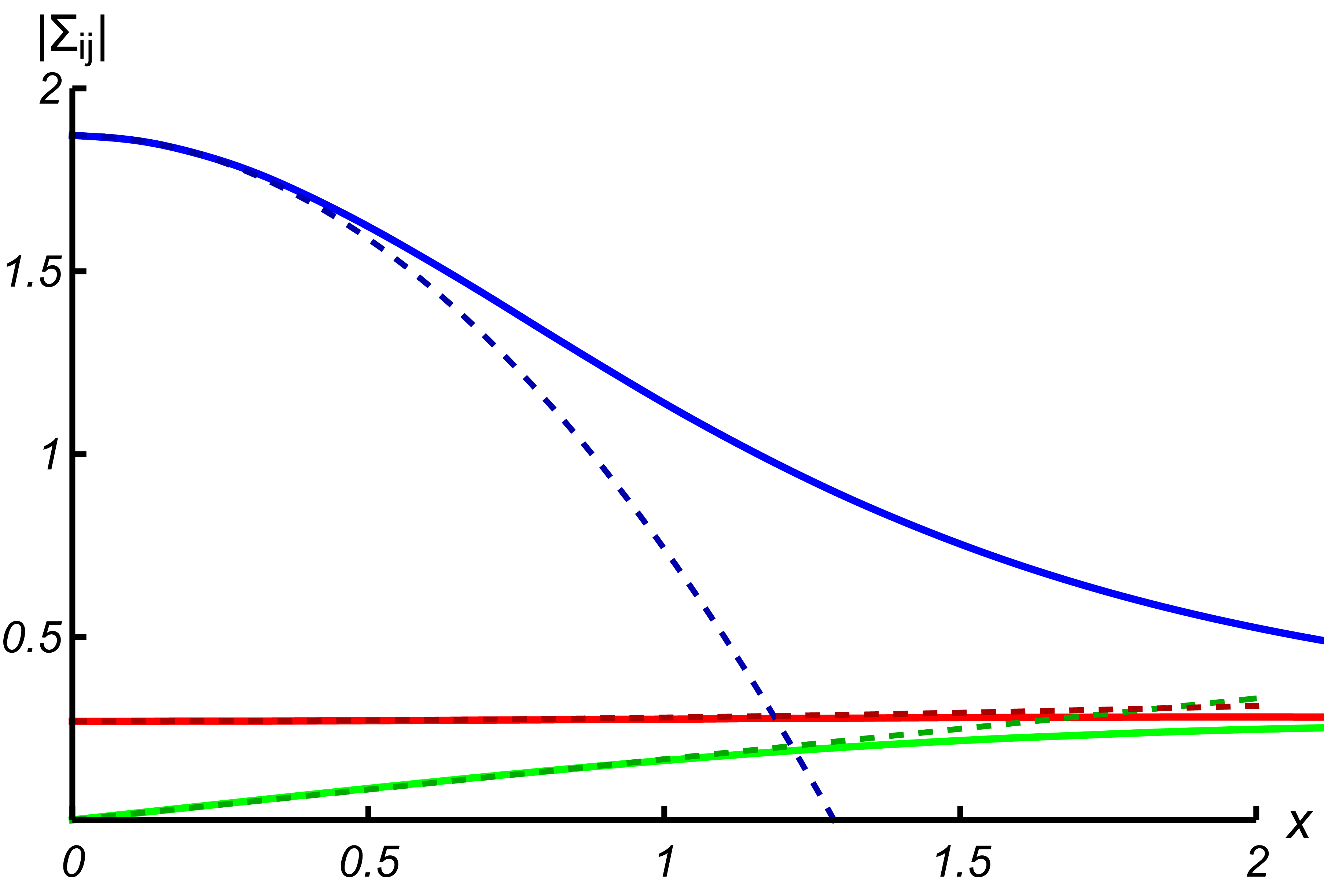}
}
\caption{Matrix elements of the dimensionless conductivity
  $\widehat\Sigma$ for small $x=\mu/T$. The blue, green, and red
  curves correspond to $\Sigma_{11}$, $\Sigma_{21}$, $|\Sigma_{22}|$,
  respectively. The dashed lines indicate the leading behavior close
  to charge neutrality.}
\label{fig5:sijdp}
\end{figure}

The leading-order correction is given by
\begin{eqnarray*}
&&
\!\!\!\!\!\delta\widehat{\textswab{M}}
=
\frac{x}{54\zeta(3)\ln2}
\\
&&
\\
&&
\times
\!
\begin{pmatrix}
- 16x\ln^22 & 27\zeta(3)\!-\!4\pi^2\!\ln2 \cr
27\zeta(3)\!-\!4\pi^2\!\ln2 & 
2\pi^2x\!
\left[\frac{\pi^2}{48\ln2}\!+\!\frac{\pi^2\!\ln2}{9\zeta(3)}\!-\!1\right]
\end{pmatrix}\!.
\end{eqnarray*}
The matrix $\widehat{\textswab{S}}_{xx}$ can be expanded in the same
way, using the expansion of the scattering rates (\ref{tausdp}):
\[
\widehat{\textswab{S}}_{xx} = \widehat{\textswab{S}}_{xx}(0) 
+ \delta \widehat{\textswab{S}}_{xx} + {\cal O}(x^3),
\]
where 
\begin{equation}
\label{sxx0}
\widehat{\textswab{S}}_{xx}(0)
\!=\!
\frac{\pi}{2T\ln2}\!
\left[\!
\begin{pmatrix}
  \tau_{11}^{-1} & 0 \cr
  0 & \tau_{22}^{-1}
\end{pmatrix}
\!+\!
\frac{1}{\tau_{\rm dis}}
\widehat{\textswab{M}}
\right]\!,
\end{equation}
and
\[
\delta \widehat{\textswab{S}}_{xx}
=
\frac{\alpha_g^2}{8\ln^22}\delta\widehat{\textswab{T}}
+
\frac{\pi}{2T\ln2}\tau_{\rm dis}^{-1}\delta\widehat{\textswab{M}},
\]
with
\[
\delta\widehat{\textswab{T}}
=
x\begin{pmatrix}
\frac{x}{t_{11}^{(2)}}\!-\!\frac{1}{8\ln2}\frac{x}{t_{11}^{(0)}} 
& 1/t_{12}^{(1)} \cr
1/t_{12}^{(1)} 
& \frac{x}{t_{22}^{(2)}}\!-\!\frac{1}{8\ln2}\frac{x}{t_{22}^{(0)}}
\end{pmatrix}.
\]
Combining the above matrices, one finds the leading corrections to the
conductivity matrix in the vicinity of the Dirac point, see
Fig.~\ref{fig5:sijdp}.

Equations (\ref{hydro}) and (\ref{djs}) reviewed in this Section
represent a close set of hydrodynamic equations describing the
electronic flows in graphene in the intermediate (``hydrodynamic'')
temperature window \cite{rev,luc}. So far, these equations were mostly
studied within linear response (nonlinear phenomena were discussed,
e.g., in Ref.~\onlinecite{hydro1}). The hydrodynamic collective modes
are also obtained by linearizing the hydrodynamic equations.

\section{Linearized hydrodynamic theory at ${\bs{B}=0}$}

In this Section, we discuss the linearization of the hydrodynamic
theory in graphene suitable for a discussion of the bulk collective
modes in the absence of the magnetic field, which is the primary focus
of this paper.

Within linear response one considers small deviations of hydrodynamic
quantities from their equilibrium values. At equilibrium, the
stationary fluid is characterized by vanishing macroscopic currents
and homogeneous thermodynamic quantities. Equilibrium quantities are
most conveniently expressed in terms of the equilibrium values of
temperature and chemical potential:
\begin{subequations}
\label{equil}
\begin{eqnarray}
\label{equilt}
&&
\mu\!=\!\bar{\mu}, \qquad T\!=\!\bar{T}, \qquad \mu_I\!=\!0, \qquad x \!=\! \bar{\mu}/\bar{T},
\\
&&
\nonumber\\
&&
n = \bar{n} = \frac{N\bar{T}^2}{2\pi v_g^2} \tilde{n}, 
\quad\quad
n_I = \bar{n}_I = \frac{N\bar{T}^2}{2\pi v_g^2} \tilde{n}_I,
\nonumber\\
&&
\nonumber\\
&&
P = \bar{P} = \frac{N\bar{T}^3}{2\pi v_g^2}\tilde{n}_E,
\quad
W=3\bar{P}, 
\quad
s=\frac{3\bar{P} - \bar{\mu} \bar{n}}{\bar{T}}.
\nonumber
\end{eqnarray}
Finally, the electric potential is homogeneous as well
\begin{equation}
\label{equile}
\varphi = \bar{\varphi}, \quad \bs{E} = - \bs{\nabla}\bar\varphi = 0,
\quad
\bs{j}=\bs{j}_I=\bs{j}_E=0.
\end{equation}
\end{subequations}
The values $\bar{\mu}$, $\bar{\varphi}$, and $\bar{T}$ are determined by the
environment in which the system is placed or, in other words, by the
boundary conditions.

Once the system is subjected to a weak external voltage and
temperature gradient, the hydrodynamic velocity $\bs{u}$ acquires a
nonzero value and thermodynamic quantities become inhomogeneous. To the
lowest (linear) order, one introduces small inhomogeneous fluctuations
of the equilibrium quantities (not all being independent)
\begin{subequations}
\label{neqqs}
\begin{equation}
\label{nech}
\mu = \bar{\mu}+\delta\mu,
\qquad
T = \bar{T}+\delta T,
\qquad
\varphi = \bar{\varphi} + \delta\varphi,
\end{equation}
\begin{equation}
\label{nen}
n = \bar{n} + \delta{n}, 
\qquad
n_I = \bar{n}_I + \delta{n}_I, 
\qquad
P = \bar{P} + \delta{P},
\end{equation}
as well as small values for those quantities that vanish in equilibrium
\begin{equation}
\label{ujni}
\bs{u}, \qquad \mu_I.
\end{equation}

The macroscopic currents have the form (\ref{jlr}). Within linear
response, the nonequilibrium corrections (\ref{jlr}) [in general given
  in Eq.~(\ref{djs})] may be expressed as
\begin{equation}
\label{djslr}
\begin{pmatrix}
\delta\bs{j} \cr
\delta\bs{j}_I
\end{pmatrix}
=\widehat\Sigma
\begin{pmatrix}
-e \bs{\nabla}\delta\zeta + x \bs{\nabla}\delta{T} \cr
-\bs{\nabla}\mu_I
\end{pmatrix}\!,
\end{equation}
where $\widehat\Sigma$ is evaluated at equilibrium and
\begin{equation}
\label{zeta}
\delta\zeta = \delta\varphi+\frac{1}{e}\delta\mu,
\end{equation}
is the electrochemical potential. Here we used the fact that $\mu_I$
and $\bs{\nabla}\delta{T}$ are both assumed to be small, so that their
products, e.g., $\mu_I\bs{\nabla}\delta{T}$, have to be neglected.

The same corrections can be expressed in terms of the density fluctuations
rather than the chemical potentials \cite{hydro1}
\begin{equation}
\label{djslr2}
\begin{pmatrix}
\delta\bs{j} \cr
\delta\bs{j}_I
\end{pmatrix}\!
\!=\widehat\Sigma\!
\begin{pmatrix}
e\bs{E} \cr
0
\end{pmatrix}
\!-
\frac{\bar T^2}{\cal T}
\widehat\Sigma'
\!\begin{pmatrix}
\bs{\nabla}\delta\tilde{n} \!-\! \frac{2\tilde{n}}{3\tilde{n}_E} \bs{\nabla}\delta\tilde{n}_E \cr
\bs{\nabla}\delta\tilde{n}_I \!-\! \frac{2\tilde{n}_I}{3\tilde{n}_E} \bs{\nabla}\delta\tilde{n}_E
\end{pmatrix}\!,
\end{equation}
with dimensionless fluctuations of the densities and pressure
[cf. Eqs.~(\ref{nech}) and (\ref{nen})] defined as
\begin{equation}
\label{nend}
\delta n = \frac{N\bar{T}^2}{2\pi v_g^2} \delta\tilde{n},
\quad
\delta n_I = \frac{N\bar{T}^2}{2\pi v_g^2} \delta\tilde{n}_I,
\quad
\delta P = \frac{N\bar{T}^3}{2\pi v_g^2} \delta\tilde{n}_E,
\end{equation}
the quantity ${\cal T}$ is related to the equilibrium compressibility
\cite{me1,me3,hydro1,hydro0}
\begin{equation}
\label{comp}
\frac{\partial\bar n}{\partial\bar\mu} = \frac{N{\cal T}}{2\pi v_g^2},
\quad
{\cal T} = 2\bar{T}\ln2\cosh\frac{\bar\mu}{2\bar T},
\end{equation}
and finally
\begin{equation}
\label{sigmap}
\widehat\Sigma'=
\widehat{\textswab{M}}\,
\widehat{\textswab{S}}_{xx}^{-1},
\qquad
\widehat\Sigma=\widehat\Sigma'\,
\widehat{\textswab{M}}.
\end{equation}
\end{subequations}
The expressions (\ref{djslr}) and (\ref{djslr2}) are completely
equivalent, however one has to be careful with the electric field.
Indeed, electrical conductivity is typically measured as a response to
the ``total'' electric field and not to the ``external electric
field.'' The total electric field includes the so-called Vlasov
self-consistency \cite{me1,luc,rev,hydro0,hydro1} taking into account
the electric field induced by the density fluctuations. The latter can
be obtained using Poisson's equation
\begin{subequations}
\label{vlasov}
\begin{equation}
\label{v3d}
\bs{E}_V = - e\bs{\nabla}\!\int\! d^2r' \frac{\delta n(\bs{r}')}{|\bs{r}\!-\!\bs{r}'|}.
\end{equation}
This relation simplifies in gated structures, where \cite{mr1,ash}
\begin{equation}
\label{vg}
\bs{E}_V = - \frac{e}{C}\bs{\nabla}\delta n(\bs{r}).
\end{equation}
\end{subequations}
Here ${C=\varepsilon/(4\pi d)}$ is the gate-to-channel capacitance per
unit area, $d$ is the distance to the gate, and $\varepsilon$ is the
dielectric constant. This approximation neglects the long-ranged
(dipole-type) part of the Coulomb interaction (screened by the gate) and
is valid as long as the charge density $n(\bs{r})$ varies on length
scales much longer than $d$.

Linearizing the hydrodynamic equations (\ref{hydro}) we find
\begin{subequations}
\label{hydrolin1}
\begin{equation}
\label{nseqlin1}
\frac{3\bar{P}}{v_g^2}\partial_t\bs{u}
+
\bs{\nabla} \delta P
=
\eta \Delta\bs{u}
+
e\bar{n}\bs{E}
-
\frac{3\bar{P}\bs{u}}{v_g^2\tau_{{\rm dis}}},
\end{equation}
\begin{equation}
\partial_t \delta{n} + \bar{n}\bs{\nabla}\!\cdot\!\bs{u} + \bs{\nabla}\!\cdot\!\delta\bs{j} = 0,
\end{equation}
\begin{equation}
\partial_t \delta{n}_I + \bar{n}_I\bs{\nabla}\!\cdot\!\bs{u} + \bs{\nabla}\!\cdot\!\delta\bs{j}_I 
= - \delta{n}_I/\tau_R,
\end{equation}
\begin{equation}
\label{tteqlin1}
2\partial_t\delta{P} + 3\bar{P}\bs{\nabla}\!\cdot\!\bs{u} = - 2\delta{P}/\tau_{RE}.
\end{equation}
\end{subequations}
Notice that the linearized ``thermal transport'' equation
(\ref{tteqlin1}) is completely equivalent (within linear response) to
the continuity equation for the energy flow; see
Refs.~\onlinecite{rev,luc,hydro0,hydro1,me1}. The energy relaxation
term in Eq.~(\ref{tteqlin1}) was derived in Ref.~\cite{meig1}.

At this point one has to choose the set of independent variables.
Based on the form of the linearized equations (\ref{hydrolin1}), one
can choose $\delta{n}$, $\delta{n}_I$, and $\delta{P}$. Together with
the two components of $\bs{u}$ one has five variables for five
differential equations (\ref{hydrolin1}). This set was used in
Ref.~\onlinecite{hydro0} to discuss collective modes in the electronic
fluid.

An alternative choice based on the form of dissipative corrections
(\ref{djslr}) may include $\delta\zeta$, $\mu_I$, and $\delta{T}$. These
variables were chosen in Ref.~\onlinecite{alf} for the discussion of
the role of the imbalance mode in thermoelectric effects. Indeed,
using the thermodynamic relation \cite{me1,luc,dau6,alf}
\begin{equation}
\label{dpdmu}
dP = n d\mu + n_I d\mu_I + s dT,
\end{equation}
in the linearized Navier-Stokes equation
(\ref{nseqlin1}), one finds
\begin{equation}
\label{nseqlin2}
\frac{3\bar{P}}{v_g^2}\!\left(\partial_t\!+\!\tau_{{\rm dis}}^{-1}\right)\bs{u}
\!=\!
\eta \Delta\bs{u}
\!-\!
e\bar{n}\bs{\nabla}\zeta
\!-\!
\bar{n}_I\bs{\nabla}\mu_I
\!-\!
\frac{3\bar{P}\!-\!\bar{n}\bar\mu}{\bar{T}}\bs{\nabla}\delta{T},
\end{equation}
where we combine the electric and chemical potential into the
electrochemical potential (\ref{zeta}). Given that the densities and
pressure are given by known functions of the chemical potentials and
temperature, see Eqs.~(\ref{equilt}), it's a matter of simple algebra
to express the rest of Eqs.~(\ref{hydrolin1}) in terms of
$\delta\zeta$, $\mu_I$, and $\delta{T}$.

While the choice of the thermodynamic variables is a matter of taste,
there is an important distinction between static and dynamic response
\cite{luc}. Static linear response equations contain only the
electrochemical potential $\zeta$. However, the dynamic part of
Eq.~(\ref{tteqlin1}) contains the chemical potential
only. Consequently, one has to be careful considering response
functions that depend on time and spatial coordinates at the same
time. In this case, an additional equation (\ref{vlasov}) describing
Vlasov self-consistency has to be taken into account \cite{hydro1}.

\section{Collective modes at ${\bs{B}=0}$}

\begin{figure*}[t]
\centerline{\includegraphics[width=0.4\textwidth]{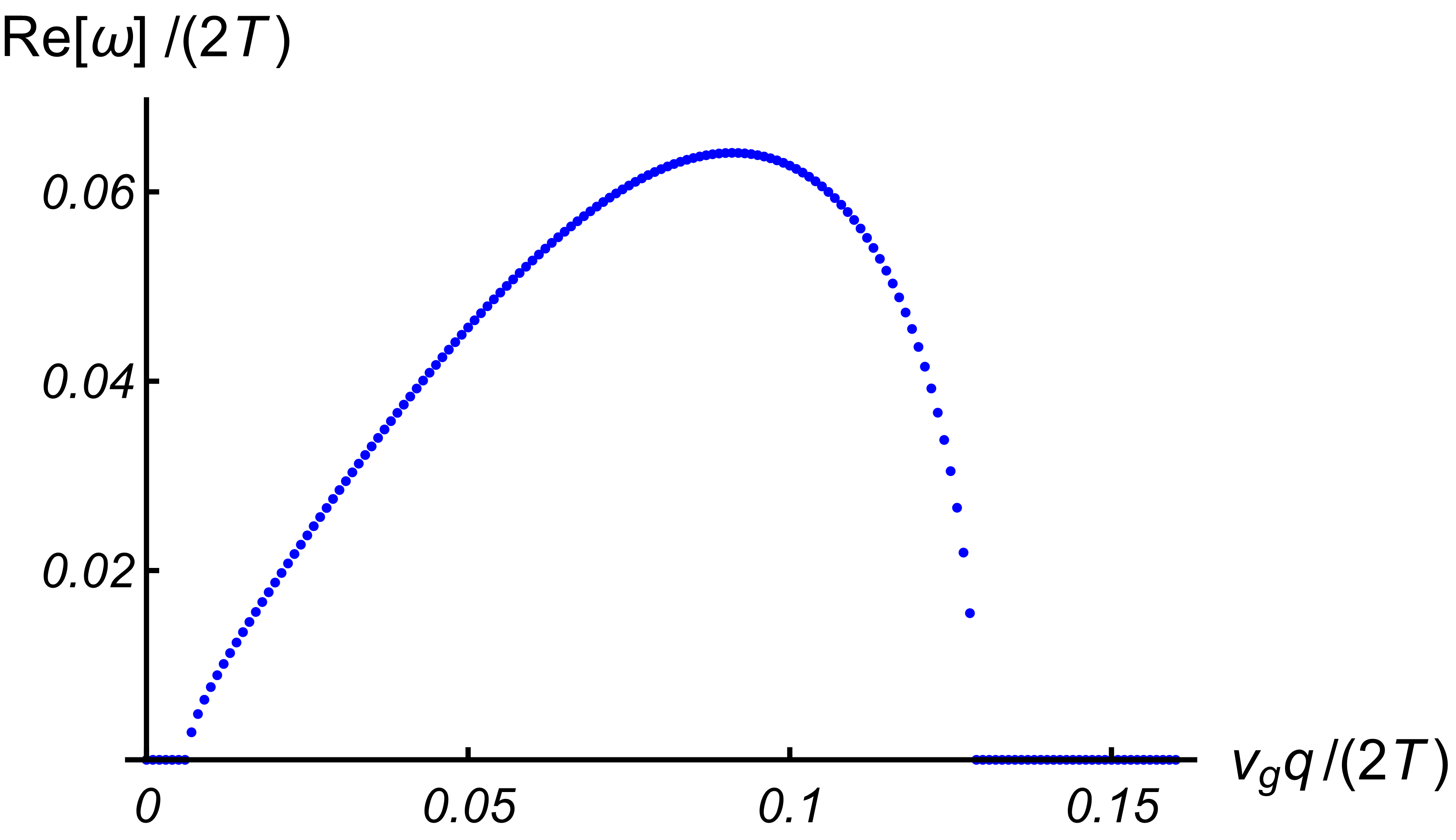}
\qquad\qquad\qquad
\includegraphics[width=0.4\textwidth]{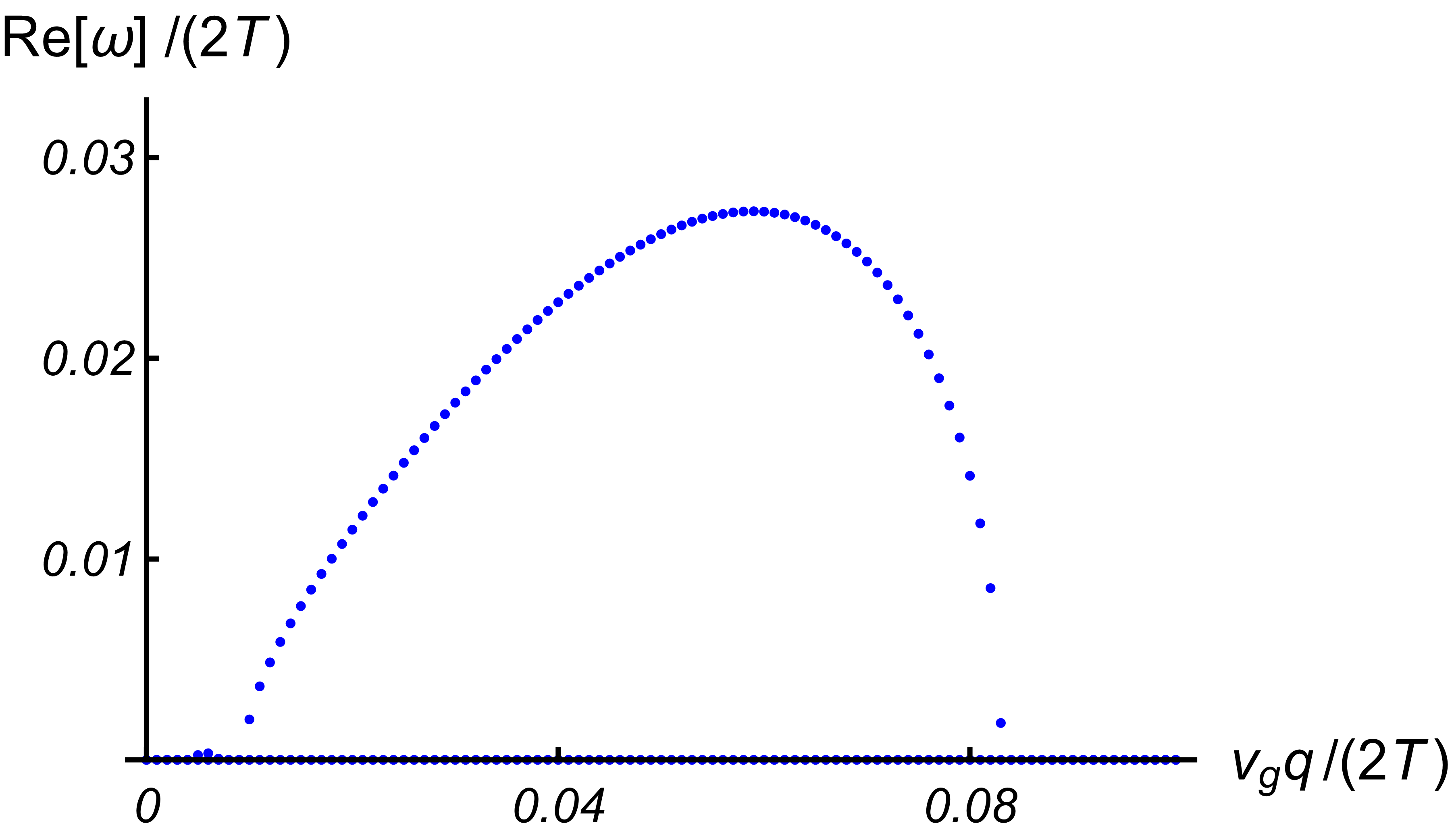}
}
\caption{Real part of the sound dispersion in moderately doped, gated
  graphene in the presence of both weak disorder and viscosity. Left
  pane: results for $n=10^{12}\,$cm$^{-2}$. Right panel: same for
  $n=10^{11}\,$cm$^{-2}$. The right panel also shows the zero mode
  Eq.~(\ref{zeromode}). }
\label{fig6:rom}
\end{figure*}

Collective modes in the electronic fluid were considered within the
same approach in Ref.~\onlinecite{hydro1}, see also
Refs.~\onlinecite{luc,mfs,mfss}. These are the eigenmodes of the
linearized equations (\ref{hydrolin1}). The most convenient choice of
variables for this task is the density-pressure variables,
$\delta{n}$, $\delta{n}_I$, and $\delta{P}$, and the velocity
$\bs{u}$. The dissipative corrections to the currents are given by
Eq.~(\ref{djslr2}) and the electric field in Eq.~(\ref{nseqlin1}) is
the total electric field.

Now, it is convenient to solve linear differential equations with the
help of the Fourier transform. Using the standard convention
\[
\bs{u}(t, \bs{r}) = 
\int \frac{d\omega d\bs{q}}{(2\pi)^3} e^{-i\omega t + i \bs{q}\bs{r}} \bs{u}(\omega, \bs{q}),
\]
we rewrite Eqs.~(\ref{hydrolin1}) in the dimensionless form
\begin{subequations}
\label{hlk8}
\begin{equation}
\label{nshlk8}
\left(\!\tilde\omega
\!+\!i\frac{1\!+\!\tilde q^2\tilde\ell_G^2}{\tilde\tau_{{\rm dis}}}\right)\tilde{n}_E\bs{\rm v}
-
\frac{1}{3}\tilde{\bs{q}} \delta \tilde{n}_E
-\tilde{\bs{q}}\tilde V_q \frac{e\tilde n}{6} \delta\tilde n
=
\frac{i}{6}e\tilde{n}\bs{\cal E}_0,
\end{equation}
\begin{equation}
\tilde\omega\delta\tilde{n} - \tilde{n}\tilde{\bs{q}}\!\cdot\!\bs{\rm v} 
- \frac{2\pi}{N} \tilde{\bs{q}}\!\cdot\!\delta\tilde{\bs{j}} = 0,
\end{equation}
\begin{equation}
\left(\!\tilde\omega\!+\!\frac{i}{\tilde\tau_{R}}\right)\delta\tilde{n}_I 
- \tilde{n}_I\tilde{\bs{q}}\!\cdot\!\bs{\rm v} - \frac{2\pi}{N}\tilde{\bs{q}}\!\cdot\!\delta\tilde{\bs{j}}_I 
= 0,
\end{equation}
\begin{equation}
\label{tteqhlk8}
\left(\!\tilde\omega\!+\!\frac{i}{\tilde\tau_{RE}}\right)\delta\tilde{n}_E 
- \frac{3}{2}\tilde{n}_E\tilde{\bs{q}}\!\cdot\!\bs{\rm v} = 0,
\end{equation}
\end{subequations}
where
\begin{subequations}
\begin{equation}
\label{dqs}
\tilde{\bs{q}} = v_g\bs{q}/(2\bar{T}),
\quad
\tilde\omega = \omega/(2\bar{T}),
\quad
\bs{\rm v} = \bs{u}/v_g,
\end{equation}
\begin{equation}
\label{dqs2}
\tilde\tau_{j} = 2\bar{T}\tau_{j}
\,\,
(j={\rm dis},\,R,\,RE),
\quad
\bs{\cal E}_0 = v_g\bs{E}_0\bar{T}^{-2},
\end{equation}
the dimensionless Gurzhi length is defined so that
\begin{equation}
\tilde{q}\tilde\ell_G = q \ell_G,
\end{equation}
and the self-consistent Vlasov potential is given by
\begin{equation}
\label{vf}
\tilde V_q = \frac{eN\bar T}{\pi v_g^2}V_{s},
\qquad
V_{s}(\bs{q}) =
\begin{cases}
e/C, & \text{gated}, \cr
2\pi e/q, & \text{Coulomb},
\end{cases}
\end{equation}
\end{subequations}
\begin{widetext}
\noindent
Finally, the dimensionless form of the dissipative corrections to the
macroscopic currents is given by
\begin{eqnarray}
\label{djslr8}
\begin{pmatrix}
\delta\tilde{\bs{j}} \cr
\delta\tilde{\bs{j}}_I
\end{pmatrix}
=\widehat\Sigma
\begin{pmatrix}
e\bs{\cal E}_0\!-\!i\tilde{\bs{q}}\tilde V_q\delta\tilde n \cr
0
\end{pmatrix}
-
\frac{i\tilde{\bs{q}}\,\widehat{\Sigma}^{\prime}}{\ln2\cosh\frac{x}{2}}
\!\begin{pmatrix}
\delta\tilde{n} \!-\! \frac{2\tilde{n}}{3\tilde{n}_E} \delta\tilde{n}_E \cr
\delta\tilde{n}_I \!-\! \frac{2\tilde{n}_I}{3\tilde{n}_E} \delta\tilde{n}_E
\end{pmatrix}\!.
\end{eqnarray}

The collective modes can now be found by analyzing the system of
Eqs.~(\ref{hlk8}). For convenience, it can be written in the
matrix form
\begin{equation}
\label{mat0}
\begin{pmatrix}
\tilde\omega \!+\! 
\frac{i2\pi \tilde{q}^2}{N}\!\!\left[\!\Sigma_{11}\tilde V_q\!+\!\frac{\Sigma'_{11}}{\ln2\cosh\frac{x}{2}}\!\right]&  
\frac{i2\pi \tilde{q}^2\Sigma'_{12}}{N\ln2\cosh\frac{x}{2}} &
-\frac{i4\pi \tilde{q}^2}{3N}\frac{\Sigma'_{11}\tilde n\!+\!\Sigma'_{12}\tilde n_I}{\tilde n_E\ln2\cosh\frac{x}{2}} &  -\tilde n \tilde{\bs{q}}\cr
\frac{i2\pi \tilde{q}^2}{N}\!\left[\Sigma_{21}\tilde V_q\!+\!\frac{\Sigma'_{21}}{\ln2\cosh\frac{x}{2}}\right]&  
\tilde\omega\!+\!\frac{i}{\tilde\tau_R}\!+\!\frac{i2\pi \tilde{q}^2\Sigma'_{22}}{N\ln2\cosh\frac{x}{2}} & 
-\frac{i4\pi \tilde{q}^2}{3N}\frac{\Sigma'_{21}\tilde n\!+\!\Sigma'_{22}\tilde n_I}{\tilde n_E\ln2\cosh\frac{x}{2}} & 
-\tilde n_I \tilde{\bs{q}} \cr
0 &  0 &  \tilde\omega\!+\!\frac{i}{\tilde\tau_{RE}} & -\frac{3}{2}\tilde n_E \tilde{\bs{q}} \cr
-\tilde{\bs{q}}\tilde V_q \frac{\tilde n}{6} & 
0 &  
-\frac{\tilde{\bs{q}}}{3} &  
\left[\!\tilde\omega\!+\!i\frac{1+\tilde{q}^2\tilde\ell_G^2}{\tilde\tau_{{\rm dis}}}\!\right]\!\tilde n_E
\end{pmatrix}
\!\!\!
\begin{pmatrix}
\delta\tilde{n} \cr
\delta\tilde{n}_I \cr
\delta\tilde{n}_E \cr
\bs{\rm v}
\end{pmatrix}
\!=\!
\begin{pmatrix}
\frac{2\pi e}{N}\Sigma_{11}\tilde{q}{\cal E}_0 \cr
\frac{2\pi e}{N}\Sigma_{21}\tilde{q}{\cal E}_0 \cr
0 \cr
\frac{i}{6}e\tilde{n}\bs{\cal E}_0 
\end{pmatrix}\!\!.
\end{equation}
Dispersion relations of the collective modes are given by the zeros of
the determinant of the matrix in the left-hand side of (\ref{mat0})
\begin{eqnarray}
\label{detfull}
&&
\left[\tilde\omega+i\frac{1+\tilde{q}^2\tilde\ell_G^2}{\tilde\tau_{{\rm dis}}}\right]\!\!
\\
&&
\nonumber\\
&&
\qquad
\times
\left\{
\left[\tilde\omega\!+\!\frac{i}{\tilde\tau_R}\!+\!\frac{i2\pi \tilde{q}^2\Sigma'_{22}}{N\ln2\cosh\frac{x}{2}}\right]\!\!
\left[\tilde\omega\!+\!\frac{i2\pi \tilde{q}^2}{N}\!\!\left(\!\Sigma_{11}\tilde V_{\tilde{\bs{q}}}\!+\!\frac{\Sigma'_{11}}{\ln2\cosh\frac{x}{2}}\right)\!\right]\!\!
\left[\!
\left(\tilde\omega\!+\!\frac{i}{\tilde\tau_{RE}}\right)\!\!
\left(\tilde\omega\!+\!i\frac{1\!+\!\tilde{q}^2\tilde\ell_G^2}{\tilde\tau_{{\rm dis}}}\right)
\!-\!\frac{\tilde{q}^2}{2}\right]
\right.
\nonumber\\
&&
\nonumber\\
&&
\qquad\qquad
-
\frac{2\pi \tilde{q}^2\Sigma'_{12}}{N\ln2\cosh\frac{x}{2}}
\left[
\frac{\tilde{q}^2\tilde V_{\tilde{\bs{q}}} \tilde n}{6\tilde n_E}
\left(\tilde n \frac{2\pi \tilde{q}^2\Sigma'_{21}}{N\ln2\cosh\frac{x}{2}}
\!+\!
\frac{\tilde n_I}{\tilde\tau_{RE}}
\!-\!
\frac{\tilde n_I}{\tilde\tau_R}\right)
\right.
\nonumber\\
&&
\nonumber\\
&&
\qquad\qquad\qquad\qquad\qquad\qquad\qquad\qquad
-
\left.
\frac{2\pi \tilde{q}^2}{N}
\!\left(\Sigma_{21}\tilde V_{\tilde{\bs{q}}}\!+\!\frac{\Sigma'_{21}}{\ln2\cosh\frac{x}{2}}\right)\!
\!
\left[\!
\left(\tilde\omega\!+\!\frac{i}{\tilde\tau_{RE}}\right)\!\!
\left(\tilde\omega\!+\!i\frac{1\!+\!\tilde{q}^2\tilde\ell_G^2}{\tilde\tau_{{\rm dis}}}\right)
\!-\!\frac{\tilde{q}^2}{2}\right]
\!
\right]
\nonumber\\
&&
\nonumber\\
&&
\qquad\qquad
\left.
-
\frac{\tilde{q}^2\tilde V_q \tilde n^2}{6\tilde n_E}
\left(\tilde\omega\!+\!\frac{i}{\tilde\tau_{RE}}\!+\!\frac{i2\pi \tilde{q}^2\Sigma'_{11}}{N\ln2\cosh\frac{x}{2}}\right)\!
\!\left(\tilde\omega\!+\!\frac{i}{\tilde\tau_R}\!+\!\frac{i2\pi \tilde{q}^2\Sigma'_{22}}{N\ln2\cosh\frac{x}{2}}\right)\!
\right\}
=0.
\nonumber
\end{eqnarray}

\end{widetext}

The first line in Eq.~(\ref{detfull}) is the factor determining the
dispersion of the transverse fluctuations of the velocity field. Under
our assumptions this mode is completely decoupled from the rest of the
system and remains diffusive for all values of the carrier
density. This might change if one considers long-range disorder
\cite{ant20}, where it was argued to induce vortical flow near charge
neutrality.

The rest of the equation is best solved numerically. In
Fig.~\ref{fig6:rom} we present the results of a numerical calculation
of the real part of the dispersion for the two values of the carrier
density, $n=10^{12}\,$cm$^{-2}$ and
$n=10^{11}\,$cm$^{-2}$. Equation~(\ref{detfull}) was solved using the
typical values of the effective coupling constant \cite{gal,sav}
$\alpha_g=0.23$, disorder scattering time \cite{gal} $\tau_{\rm
  dis}^{-1}=1\,$THz, kinematic viscosity \cite{me2,geim1}
$\nu=0.2\,$m$^2$/s, and temperature $T=300\,$K. The result is
qualitatively similar to that shown in Fig.~\ref{fig1:sd0v}, therefore
we postpone the discussion until after we have considered the two
limiting cases where the dispersion can be obtained analytically, see
Eq.~(\ref{csv}).

\subsubsection{Collective modes in neutral graphene}
\label{hdp}

At charge neutrality, the linearized equations (\ref{hlk8}) can be
simplified using the fact that the ``conductivity matrices''
$\widehat\Sigma$ and $\widehat\Sigma'$ are block-diagonal (here we
take into account weak disorder)
\begin{equation}
\label{cmn0}
\widehat\Sigma \!= \!
\frac{1}{e^2}\!
\begin{pmatrix}
\sigma_0 & 0 \cr
0 & \sigma_{I}\delta_I 
\end{pmatrix}\!\!,
\quad
\widehat\Sigma' \!=\!
\frac{1}{e^2}\!
\begin{pmatrix}
\sigma_0 & 0 \cr
0 & \sigma_{I} 
\end{pmatrix}\!\!.
\end{equation}
As a result, the dissipative corrections (\ref{djslr8}) simplify
\begin{subequations}
\begin{equation}
\label{djdp0}
\delta\tilde{\bs{j}} = \frac{1}{e} \sigma_0 \bs{\cal E}
-\frac{i\tilde{\bs{q}}\sigma_0}{e^2\ln2}
\delta\tilde{n},
\end{equation}
\begin{equation}
\delta\tilde{\bs{j}}_I = 
-\frac{i\tilde{\bs{q}}\sigma_I}{e^2\ln2}
\left(\delta\tilde{n}_I\!-\! \frac{2\pi^2}{27\zeta(3)} \delta\tilde{n}_E\right).
\end{equation}
\end{subequations}
Using the explicit form of the equilibrium quantities \cite{me1}, we rewrite
Eqs.~(\ref{hlk8}) in the form
\begin{subequations}
\label{hlk3}
\begin{equation}
\label{nshlk3}
\tilde{\bs{q}} \delta \tilde{n}_E
-
\frac{9\zeta(3)}{2}\!\left(\!\tilde{\omega}\!+\!i\frac{1+\tilde{q}^2\tilde\ell_G^2}{\tilde\tau_{{\rm dis}}}\right)
\bs{\rm v}
=0,
\end{equation}
\begin{equation}
\label{eqkdn3}
\left(\tilde{\omega}
+i\frac{2\pi \tilde{q}^2\sigma_0}{e^2N\ln2}\right)\!\delta\tilde{n}
= \frac{2\pi\sigma_0}{eN} \tilde{\bs{q}}\!\cdot\!\bs{\cal E},
\end{equation}
\begin{equation}
\label{eqkdni3}
\left(\!\tilde{\omega}\!+\!\frac{i}{\tilde\tau_{R}}\!+\!\frac{i2\pi \tilde{q}^2\sigma_I}{e^2N\ln2}\right)\!\delta\tilde{n}_I 
\!-\! \frac{\pi^2}{6}\tilde{\bs{q}}\!\cdot\!\bs{\rm v} 
\!-\!
\frac{i4\pi^3 \tilde{q}^2\sigma_I\delta\tilde{n}_E}{27\zeta(3)Ne^2\ln2} 
\!=\! 0,
\end{equation}
\begin{equation}
\label{tteqhlk3}
2\left(\!\tilde{\omega}\!+\!\frac{i}{\tilde\tau_{RE}}\right)\!\delta\tilde{n}_E 
- \frac{9\zeta(3)}{2}\tilde{\bs{q}}\!\cdot\!\bs{\rm v} = 0.
\end{equation}
\end{subequations}
Combining Eqs.~(\ref{nshlk3}) and (\ref{tteqhlk3}) to exclude the
velocity, one finds
\[
\tilde{q}^2 \delta \tilde{n}_E
=
2\left(\!\tilde{\omega}\!+\!i\frac{1+\tilde{q}^2\tilde\ell_G^2}{\tilde\tau_{{\rm dis}}}\right)\!\!
\left(\!\tilde{\omega}\!+\!\frac{i}{\tilde\tau_{RE}}\right)\!\delta\tilde{n}_E,
\]
yielding the spectrum (\ref{csv}) [in dimensionless units; in
  Eq.~(\ref{csv}) we have neglected weak energy relaxation]
\begin{equation}
\label{cm1}
\tilde\omega = 
\sqrt{\frac{\tilde q^2}{2} 
\!- \!
\frac{1}{4}\!\left[
\frac{1\!+\!\tilde q^2\tilde\ell_G^2}{\tilde\tau_{{\rm dis}}}
\!-\!\frac{1}{\tilde\tau_{RE}}
\right]^2}
- i \frac{1\!+\!\tilde q^2\tilde\ell_G^2}{2\tilde\tau_{{\rm dis}}}-\frac{i}{2\tilde\tau_{RE}}.
\end{equation}
In the absence of dissipation this is the so-called ``cosmic sound'' wave
\cite{luc,hydro1,cosmic} with the linear dispersion (\ref{cs0}).

Same conclusions can be reached using the general form
Eq.~(\ref{detfull}). At charge neutrality, Eq.~(\ref{detfull})
factorizes
\begin{eqnarray}
\label{detdp}
&&
\left[\!\left(\!\tilde{\omega}\!+\!i\frac{1+\tilde{q}^2\tilde\ell_G^2}{\tilde\tau_{{\rm dis}}}\right)\!\!
\left(\!\tilde{\omega}\!+\!\frac{i}{\tilde\tau_{RE}}\right)\!-\!\frac{\tilde{q}^2}{2}\right]\!\!
\left[\tilde{\omega}+i\frac{1+\tilde{q}^2\tilde\ell_G^2}{\tilde\tau_{{\rm dis}}}\right]
\\
&&
\nonumber\\
&&
\quad
\times
\left[\tilde{\omega}+\frac{i}{\tilde\tau_R}+\frac{2\pi i \tilde{q}^2 \sigma_I}{Ne^2\ln2}\right]\!\!
\left[\tilde{\omega}\!+\!\frac{2\pi i \tilde{q}^2\sigma_0}{Ne^2} \left(\tilde V_{\tilde{\bs{q}}} \!+\!\frac{1}{\ln2}\right)\right]
=0.
\nonumber
\end{eqnarray}
Here the first factor yields the spectrum (\ref{cm1}), the last factor
describes the transverse fluctuations of the velocity field, while the
remaining two correspond to the charge and imbalance modes.

The sound mode (\ref{cm1}) is the energy wave not involving charge
density fluctuations [since neither Eq.~(\ref{nshlk3}) nor
  Eq.~(\ref{tteqhlk3}) contains $\delta\tilde{n}$]. Consequently, the
sound spectrum is not affected by the Vlasov self-consistency
(\ref{vlasov}).

Other modes are diffusive. Since Eqs.~(\ref{nshlk3}) and
(\ref{tteqhlk3}) are independent of the density fluctuations
$\delta\tilde{n}$ and $\delta\tilde{n}_I$, the diffusive modes can
be read off Eqs.~(\ref{eqkdn3}) and (\ref{eqkdni3}).

\begin{figure*}[t]
\centerline{\includegraphics[width=0.4\textwidth]{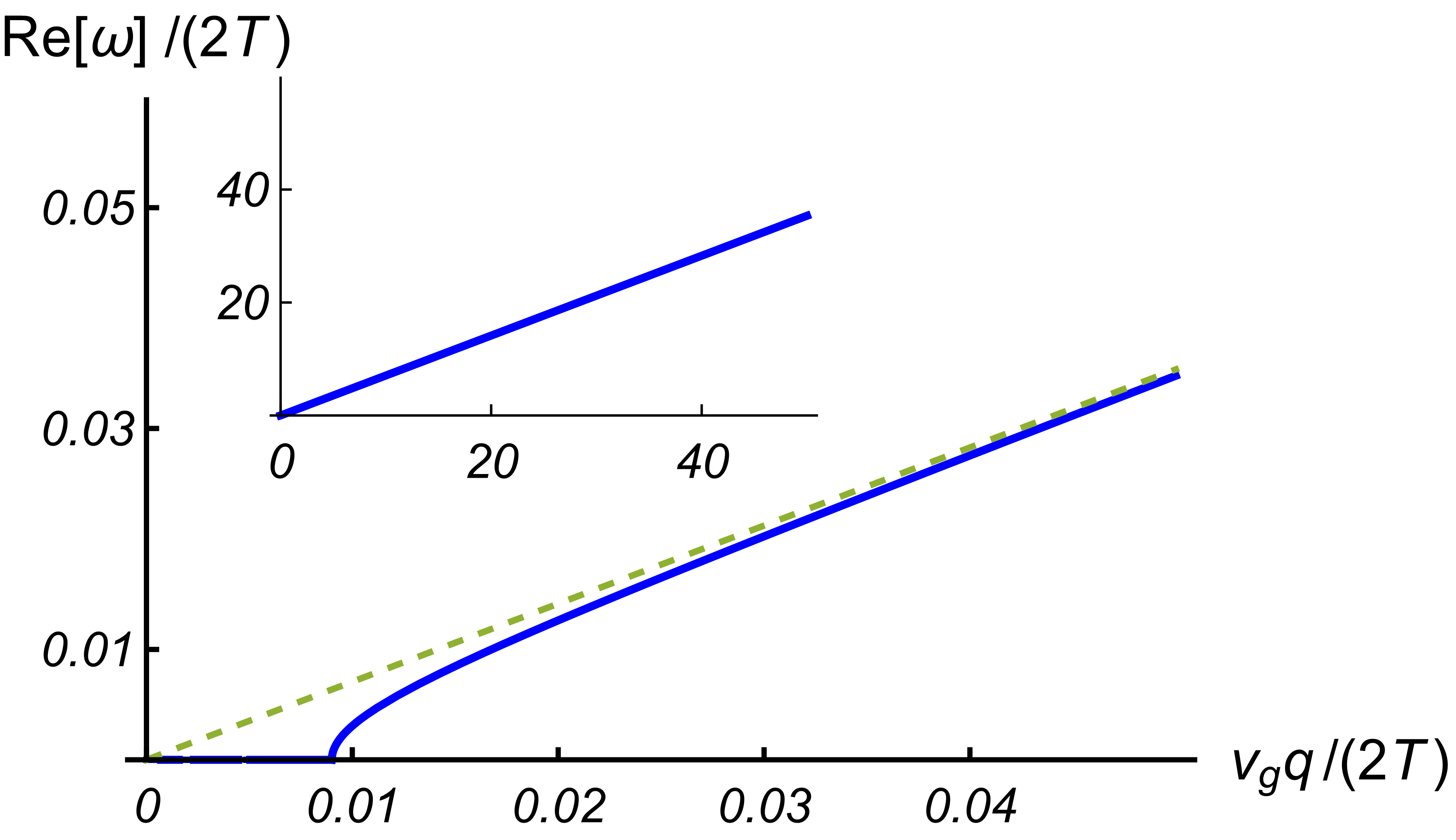}
\qquad\qquad\qquad
\includegraphics[width=0.4\textwidth]{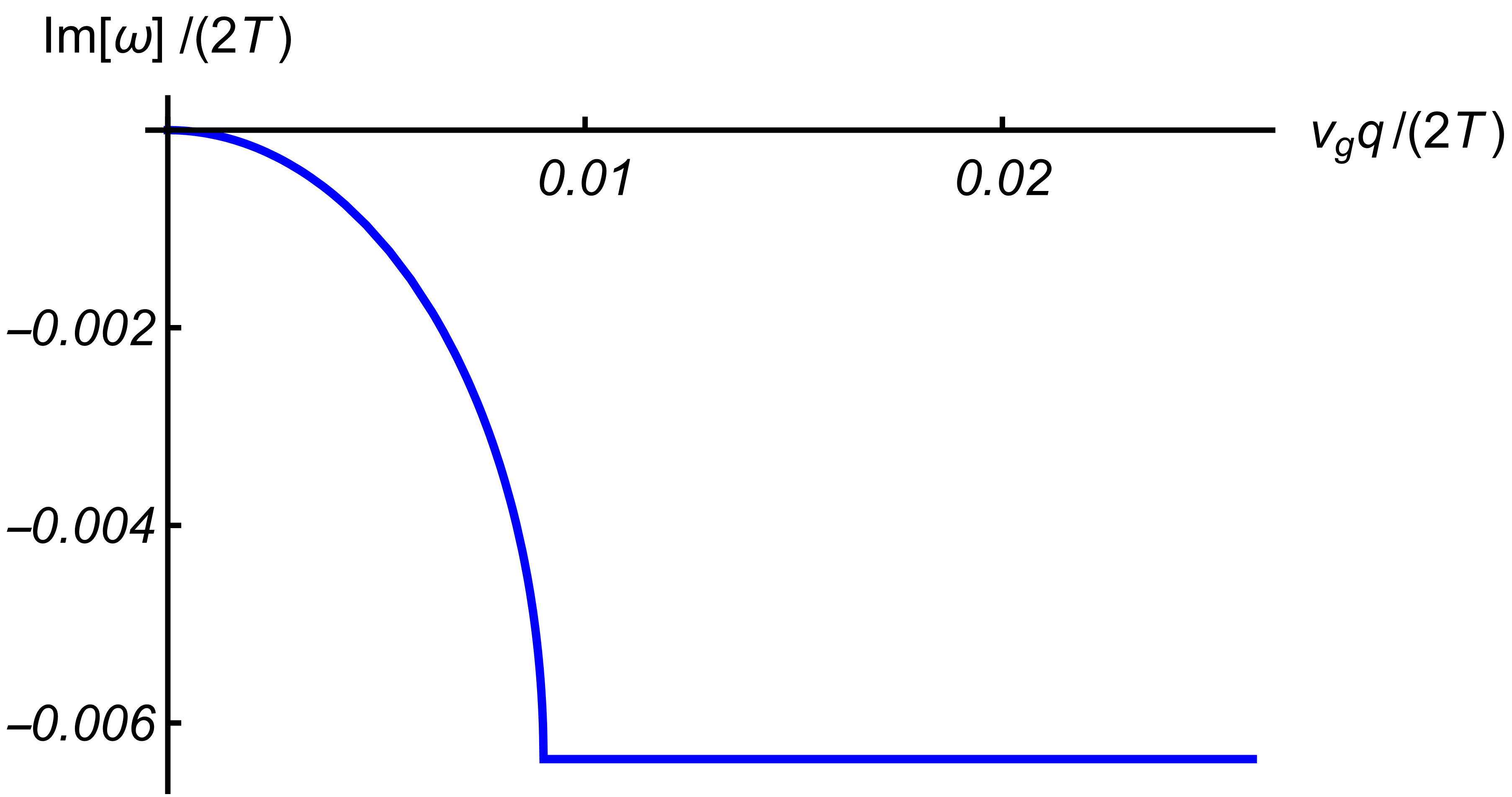}
}
\caption{Real (left panel) and imaginary (right panel) parts of the
  sound dispersion in neutral graphene neglecting viscosity. The
  dashed line represents the ideal ``cosmic sound'' dispersion
  (\ref{cs0}).}
\label{fig7:sd0}
\end{figure*}

The electric charge density fluctuations are decoupled from the rest
of the variables. Restoring the dimensionfull units and using the
explicit form (\ref{sq}) of the conductivity at charge neutrality
\cite{me1,hydro0,rev,luc,kash,mfss,mfs,schutt,das}
we can write the corresponding dispersion as
\begin{equation}
\label{cmdp}
\omega =-iD_0q^2\!\left[1\!+\!eV_{s}(q)\frac{\partial n}{\partial\mu}\right]\!,
\quad
D_0 = \frac{1}{2}\frac{v_g^2\tau_{11}\tau_{\rm dis}}{\tau_{11}\!+\!\tau_{\rm dis}}.
\end{equation}
In a gated structure the mode is diffusive with the diffusive
coefficient containing a correction due to the Vlasov
self-consistency. In the case of the long-range Coulomb interaction
the dispersion is still purely imaginary, with $\omega\sim iq$ at
small $q$.

Similarly, the imbalance mode is characterized by the diffusive spectrum
\begin{equation}
\label{imdp}
\omega =-iD_Iq^2-\frac{i}{\tau_R},
\quad
D_I = \frac{1}{2}\frac{v_g^2\tau_{22}\tau_{\rm dis}\delta_I}{\tau_{22}\delta_I\!+\!\tau_{\rm dis}},
\end{equation}
which is gapped by the recombination processes.

The hydrodynamic theory outlined in Section~\ref{ht} is justified by
the gradient expansion and hence for momenta smaller than a certain
scale defined by the electron-electron interaction
\[
q\ell_{\rm hydro}\ll1, \quad
\ell_{\rm hydro}\sim\frac{v_g}{\alpha_g^2\bar{T}}.
\]
Assuming an ultra-clean sample with $\tau_{\rm dis}\rightarrow\infty$
(where energy relaxation due to supercollisions \cite{meig1} may be
neglected, $\tau_{RE}\gg\tau_{\rm dis}$), the expression under the
square root in Eq.~(\ref{cm1}) yields
\[
\frac{v_g^2q^2}{2} - \frac{\left(1\!+\!q^2\ell_G^2\right)^2}{4\tau^2_{{\rm dis}}}
\rightarrow
\frac{v_g^2q^2}{2}\left[1 \!-\! A q^2 \ell_{\rm hydro}^2
\!- {\cal O}(\tau_{\rm dis}^{-1})\right]\!,
\]
where $A$ is a numerical coefficient. As a result, within the region
of applicability of the hydrodynamic theory the viscous term should be
neglected. The resulting dispersion acquires a simple form
\cite{hydro1}
\begin{equation}
\label{cm2}
\omega = 
\sqrt{\frac{v_g^2q^2}{2} - \frac{1}{4\tau^2_{{\rm dis}}}}
- \frac{i}{2\tau_{{\rm dis}}},
\end{equation}
illustrated in Fig.~\ref{fig7:sd0}. Now, keeping the viscous term to
the leading order, but neglecting disorder scattering \cite{svin}
yields an expansion
\begin{equation}
\label{antisv}
\omega = 
\frac{v_gq}{\sqrt{2}}\left(1 - \frac{\nu^2q^2}{4v_g^2}\right) - \frac{i\nu q^2}{2}.
\end{equation}
Similar expression was obtained in Ref.~\onlinecite{svin} based on the
phenomenological collision integral (which did not take into account
graphene-specific collinear scattering singularity). However, the
viscosity-induced correction to the real part was positive indicating
a tendency towards an indefinite growth of the dispersion instead of
the decrease towards zero implied in Eqs.~(\ref{cs0}) and (\ref{cm1})
and illustrated in Figs.~\ref{fig1:sd0v} and \ref{fig6:rom}. The sign
of the correction in Eq.~(\ref{antisv}) is, in fact, dictated by the
dissipative nature of viscosity, which represents an additional decay
mechanism and hence affects the dispersion similarly to weak disorder;
see Eq.~(\ref{cm2}). Indeed, both terms, $\tau_{\rm dis}^{-1}$ and
${\nu{q}^2}$, enter the dispersion equation [following from the first
  term in Eq.~(\ref{detdp})] on equal footing.

As shown in Refs.~\onlinecite{hydro0,hydro1,me3,drag2} the linearized
theory (\ref{hydrolin1}) has a wider applicability range due to the
kinematic peculiarity of the Dirac fermions in graphene known as the
``collinear scattering singularity'' \cite{hydro1,mfss,rev,luc}. In
the weak coupling limit, the linear response theory is valid at much
shorter length scales
\begin{equation}
\label{scasep}
q\ell_{\rm coll}\ll1, \quad
\ell_{\rm coll}\sim\frac{v_g}{\alpha_g^2\bar{T}|\ln\alpha_g|}\ll\ell_{\rm hydro}.
\end{equation}
At the same time, the viscous term is the result of the gradient
expansion that is justified at smaller momenta
\[
q\ell_{\rm hydro}\ll1,
\]
which formally restricts us to small values of $\nu q/v_g$, such that
the result (\ref{cm1}) should be expressed in terms of the expansion
(\ref{antisv}). Moreover, the imaginary part of the sound dispersion
becomes comparable to the real part at $q\ell_{\rm hydro}\sim1$, such
that the decline of the dispersion at larger $q$ shown in
Figs.~\ref{fig1:sd0v} and \ref{fig6:rom} is unlikely to be observable
anyway. Nevertheless in Figs.~\ref{fig1:sd0v}, \ref{fig6:rom}, and
\ref{fig10:sdcv} we show the sound dispersion in the whole range of
momenta to illustrate the analytic structure of our results.

For realistic model parameters, the dispersion (\ref{cm1}) shown in
Figs.~\ref{fig1:sd0v} and \ref{fig6:rom} is overdamped practically over
the whole range of momenta. In the limit of large $\tau_{\rm dis}$ and
small viscosity, the dispersion (\ref{cm1}) approaches the ideal sound
dispersion (\ref{cs0}) if
\[
(v_g\tau_{\rm dis})^{-1}\ll q\ll\ell_G^{-1}.
\]
However, taking into account the numerical prefactors and realistic
parameter values leads to Figs.~\ref{fig1:sd0v} and \ref{fig6:rom}, where
the dispersion strongly deviates from Eq.~(\ref{cs0}).

\begin{figure*}[t]
\centerline{\includegraphics[width=0.4\textwidth]{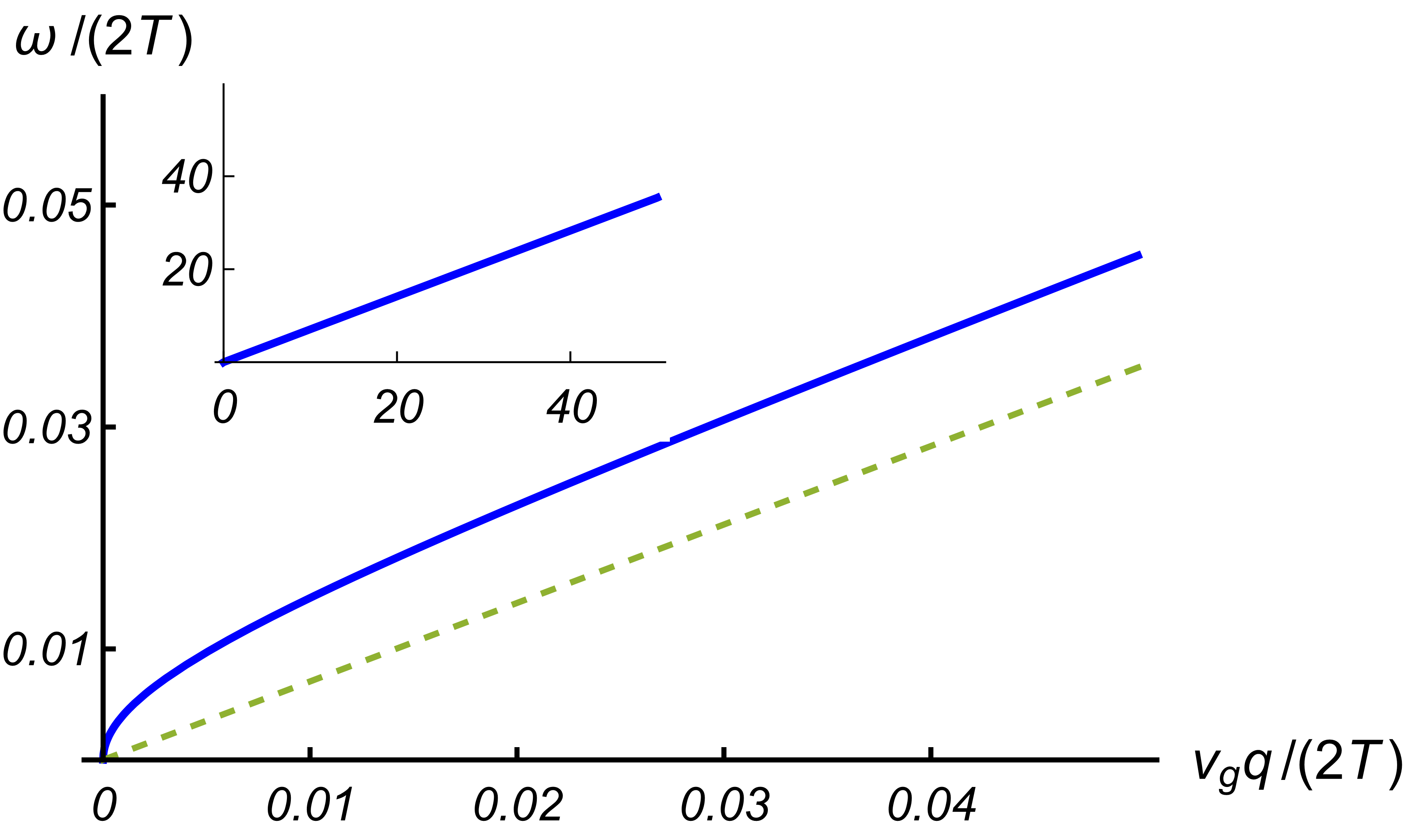}
\qquad\qquad\qquad
\includegraphics[width=0.4\textwidth]{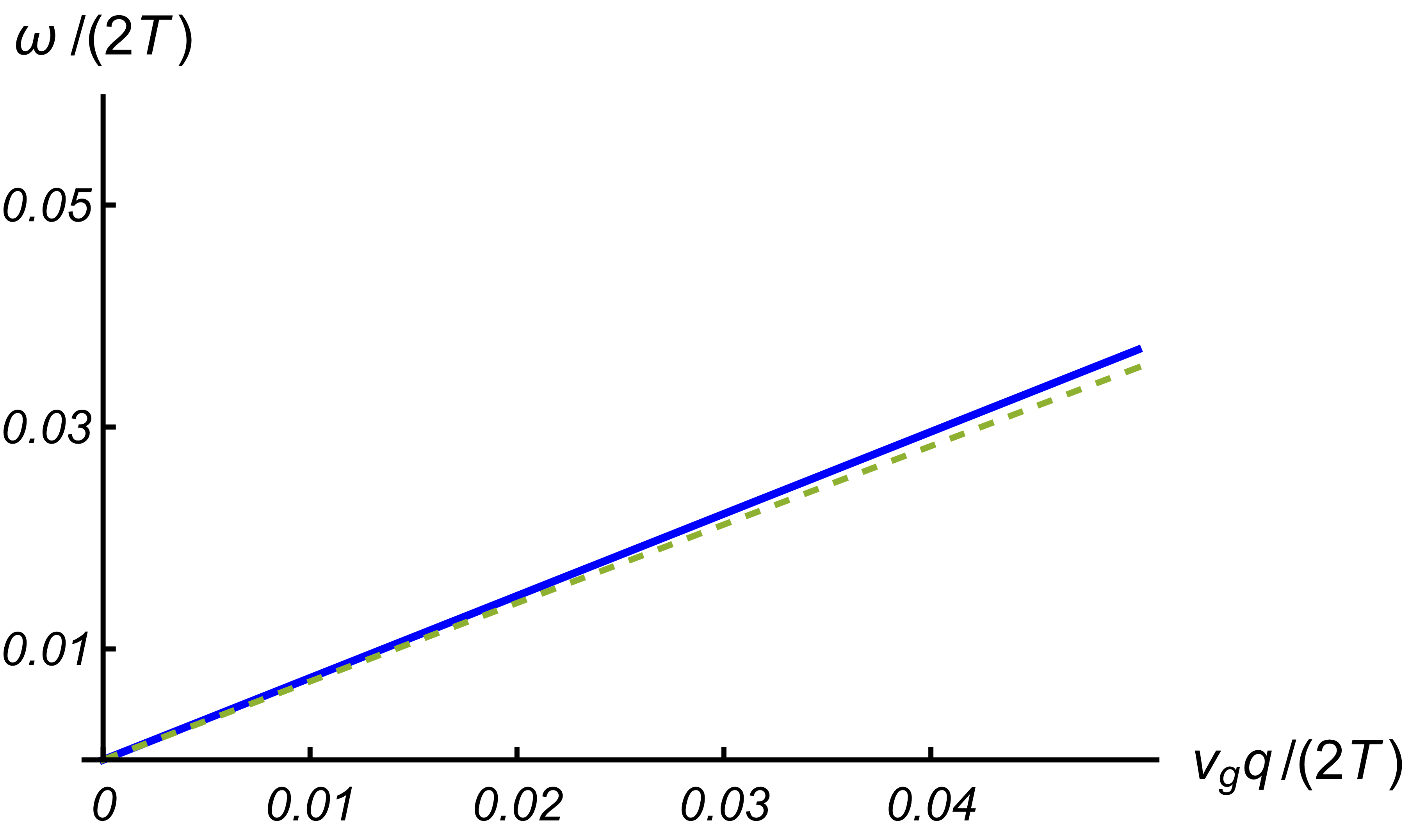}
}
\caption{Sound dispersion in strongly doped graphene neglecting both
  weak disorder and viscosity. Left panel: the result for the Coulomb
  screening, resembling the 2D plasmon for very low $q$. Right panel:
  same for a gated structure. The dashed line represents the ideal
  ``cosmic sound'' dispersion (\ref{cs0}).  }
\label{fig8:sdc0}
\end{figure*}

\subsubsection{Collective modes in the degenerate regime}

In the opposite limit of the degenerate regime, ${\mu\gg{T}}$, the matrix
in the left-hand side of Eq.~(\ref{mat0}) simplifies to
\begin{equation}
\label{matfl}
\begin{pmatrix}
\tilde\omega &  
0 &
0 &  -\tilde n \bs{q}\cr
0 &  
\tilde\omega\!+\!\frac{i}{\tilde\tau_R} & 
0 & 
-\tilde n_I \bs{q} \cr
0 &  0 & \tilde\omega\!+\!\frac{i}{\tilde\tau_{RE}} & -\frac{3}{2}\tilde n_E \bs{q} \cr
-\bs{q}\tilde V_q \frac{e\tilde n}{6} & 
0 &  
-\frac{\bs{q}}{3} &  
\left[\tilde\omega\!+\!i\frac{1+q^2\tilde\ell_G^2}{\tilde\tau_{{\rm dis}}}\right]\!\tilde n_E
\end{pmatrix}
\!,
\end{equation}
such that Eq.~(\ref{detfull}) factorizes again
\begin{widetext}
\begin{eqnarray}
\label{detfl}
\left\{\tilde\omega\left[\left(\!\tilde{\omega}\!+\!i\frac{1+\tilde{q}^2\tilde\ell_G^2}{\tilde\tau_{{\rm dis}}}\right)\!\!
\left(\!\tilde{\omega}\!+\!\frac{i}{\tilde\tau_{RE}}\right)\!-\!\frac{\tilde{q}^2}{2}\right]
-\frac{\tilde{q}^2\tilde V_qe\tilde n^2}{6\tilde n_E}\left(\!\tilde{\omega}\!+\!\frac{i}{\tilde\tau_{RE}}\right)
\right\}
\left[\tilde\omega+\frac{i}{\tilde\tau_R}\right]
\left[\tilde\omega+i\frac{1+\tilde{q}^2\tilde\ell_G^2}{\tilde\tau_{{\rm dis}}}\right]
=0.
\end{eqnarray}
\end{widetext}
The transverse velocity fluctuations remain decoupled with the same
diffusive dispersion. The imbalance mode is no longer diffusive: if
created, any imbalance density fluctuations decay exponentially in
agreement with physical intuition.

The charge and energy densities are now coupled by the self-consistent
Vlasov field. The corresponding dispersion can be found by equating
the expression in curly brackets in Eq.~(\ref{detfl}) to zero. This
leads to a cubic equation that can be solved exactly, but the analytic
solution is cumbersome and not physically transparent. Instead, we
focus on the limit ${\tau_{RE}\gg\tau_{\rm dis}}$ solving the equation
perturbatively. Neglecting energy relaxation yields two modes, one
being a flat zero mode and another the ``sound mode'' (\ref{cm1})
renormalized by the Vlasov self-consistency. To the leading order in
energy relaxation, the zero mode in a gated structure acquires the
diffusive dispersion
\begin{equation}
\label{zeromode}
\omega=-\frac{i}{\tau_{RE}}
\frac{\varkappa v_g^2q^2}{(\varkappa\!+\!2\pi C)v_g^2q^2\!+\!4\pi C\tau_{RE}^{-1}\tau_{\rm dis}^{-1}},
\end{equation}
where the Thomas-Fermi screening length is given by
\begin{equation}
\label{tfs}
\varkappa = N \alpha_g k_F = Ne^2\mu/v_g^2.
\end{equation}
In the case of the long-range Coulomb interaction, the factor $2\pi C$
should be replaced with the momentum $q$. Physically,
Eq.~(\ref{zeromode}) describes energy diffusion appearing due to
Vlasov self-consistency that couples charge and energy fluctuations.

Similarly to the above limit of neutral graphene, these results can be
obtained from a direct analysis of the linearized hydrodynamic
equations (\ref{hlk8}). In the degenerate regime ($\mu\gg T$ or
$x\gg1$), Eqs.~(\ref{hlk8}) can be simplified by noticing
that only one band contributes. For electron doping, $n\approx n_I$,
while the dissipative corrections to the currents vanish \cite{me1}
\[
\delta\bs{j}(T\ll\mu)=\delta\bs{j}_I(T\ll\mu)=0.
\]
As a result, one of Eqs.~(\ref{hlk8}) is redundant.

Assuming a gated structure and substituting the explicit form of
equilibrium densities, we find
\begin{subequations}
\label{hlk5}
\begin{equation}
\label{nshlk5}
\left[
\!\left(\!\tilde\omega\!+\!\frac{i}{\tilde\tau_{{\rm dis}}}\right)\!
\frac{x^3}{2}
\!+\! i\tilde{q}^2 \tilde\eta
\right]\!\bs{\rm v}
-
\tilde{\bs{q}} \delta \tilde{n}_E=
ie\!\left[\bs{\cal E}_0\!-\!\frac{i\tilde{\bs{q}}e}{\tilde C}\delta\tilde{n}\right]\!
\frac{x^2}{4}\!,
\end{equation}
\begin{equation}
\tilde\omega\delta\tilde{n} - (x^2/2)\tilde{\bs{q}}\!\cdot\!\bs{\rm v} = 0,
\end{equation}
\begin{equation}
\label{tteqhlk5}
2\left(\!\tilde\omega\!+\!\frac{i}{\tilde\tau_{RE}}\right)\delta\tilde{n}_E 
- (x^3/2)\tilde{\bs{q}}\!\cdot\!\bs{\rm v} = 0.
\end{equation}
\end{subequations}
Combining Eqs.~(\ref{nshlk5}) and (\ref{tteqhlk5}) one finds the
cosmic sound mode \cite{luc,hydro1,cosmic} damped by disorder and
viscosity (back to dimensionful units and for $\tau_{RE}\gg\tau_{\rm dis}$)
\begin{equation}
\label{csflg}
\omega
\!=\!
\sqrt{\frac{v_g^2q^2}{2}
\!\left[1\!+\!\frac{\varkappa}{2\pi C}\!\right]
\!-\!\frac{\left(1
\!+\!\ell_G^2 q^2\right)^{\!2}}{4\tau^2_{{\rm dis}}}}
-
\frac{i\left(1
\!+\!\ell_G^2 q^2\right)^{\!2}}{2\tau_{{\rm dis}}}\!.
\end{equation}
This is clearly the same mode as Eq.~(\ref{cm1}), albeit with the
velocity renormalized by the capacitive screening.

\begin{figure*}[t]
\centerline{\includegraphics[width=0.4\textwidth]{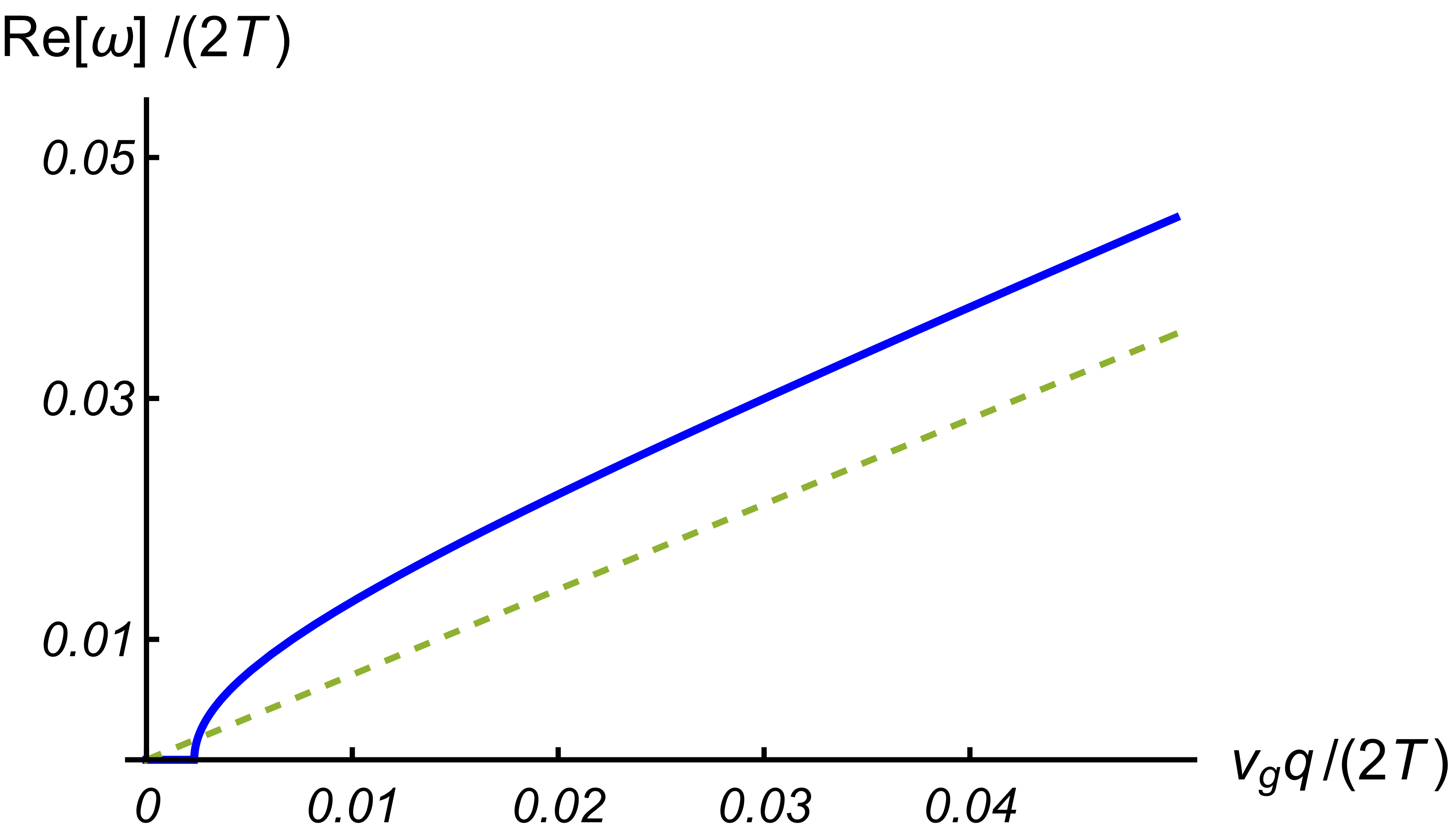}
\qquad\qquad\qquad
\includegraphics[width=0.4\textwidth]{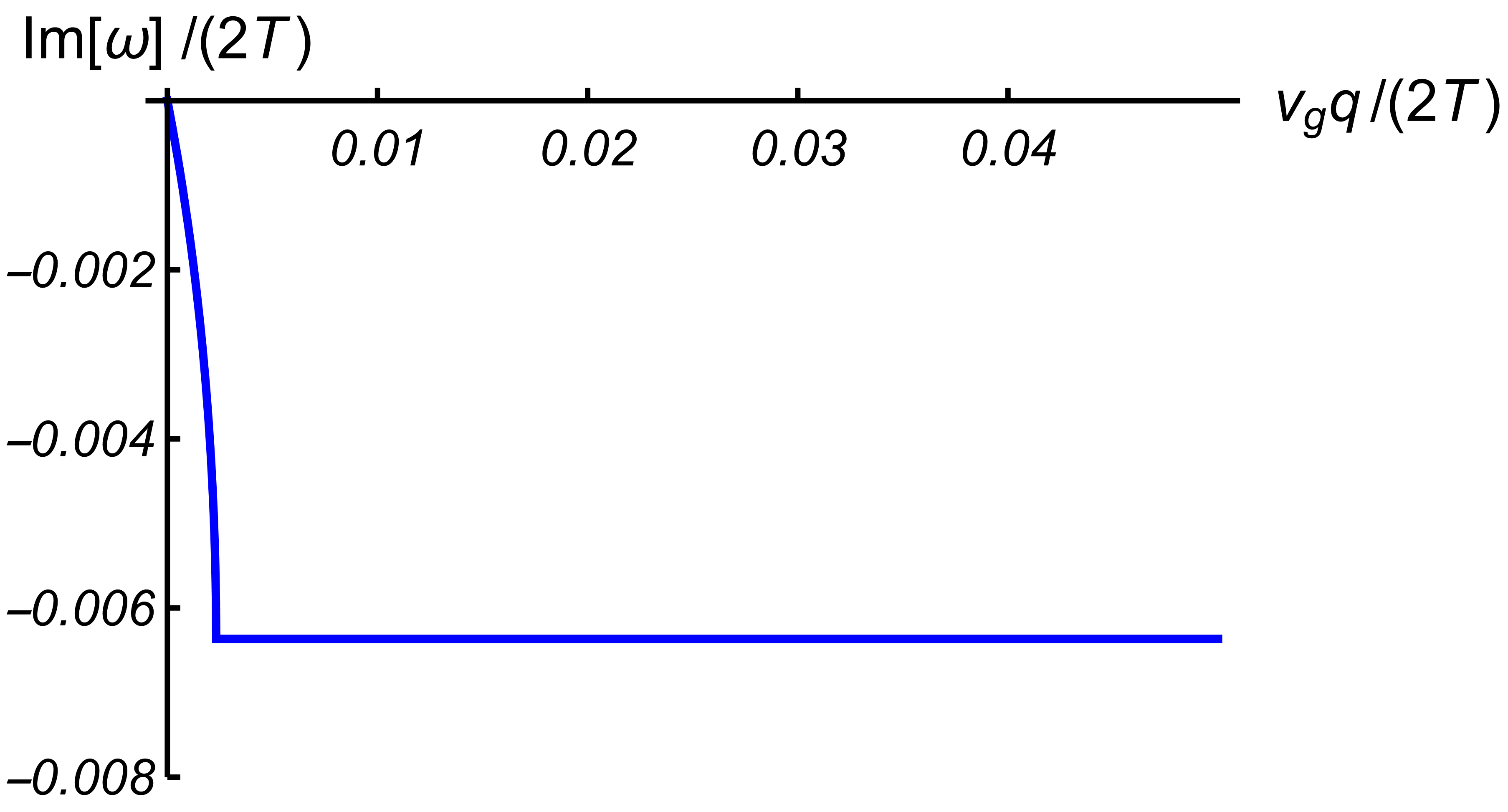}
}
\bigskip
\centerline{\includegraphics[width=0.4\textwidth]{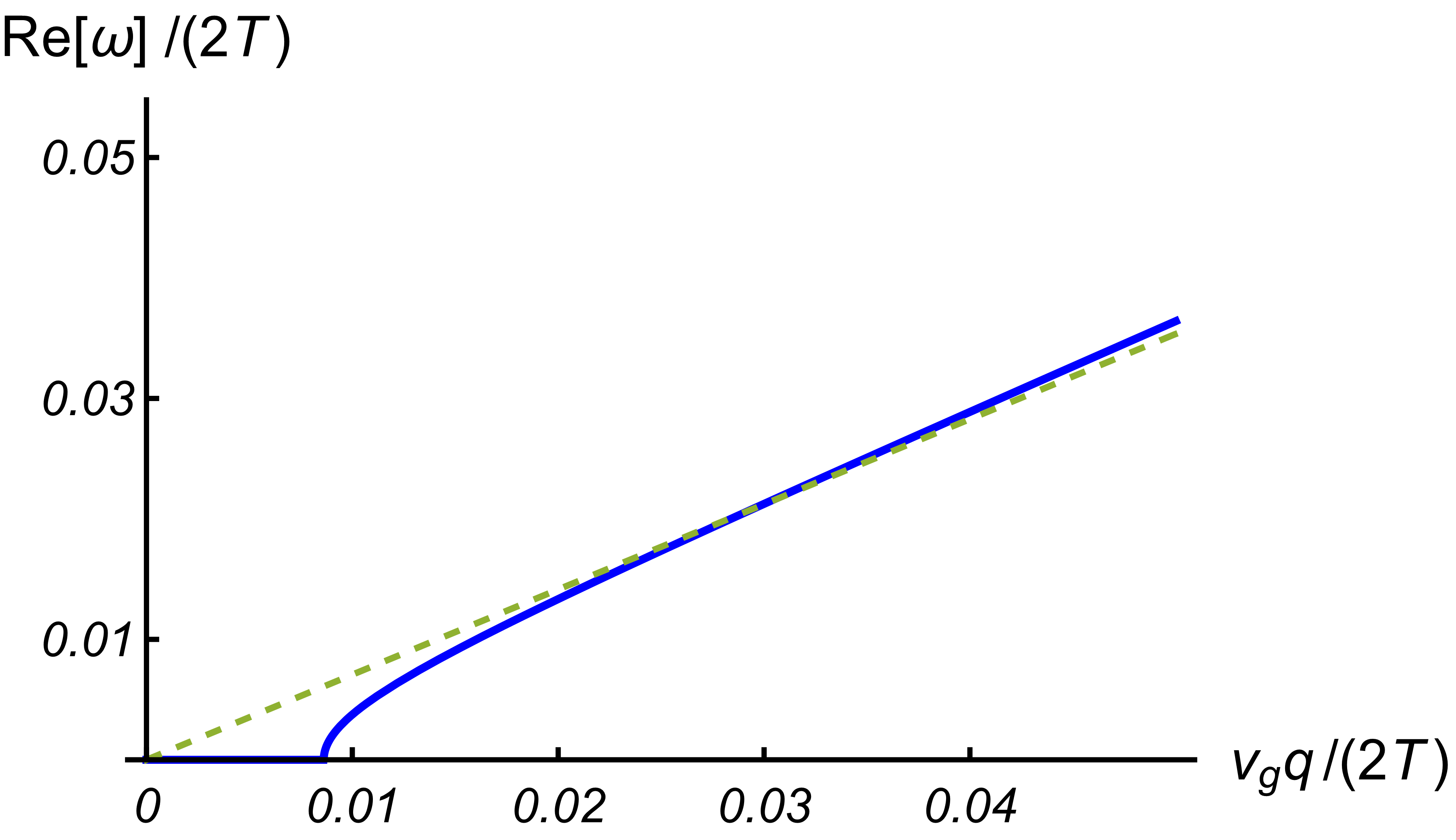}
\qquad\qquad\qquad
\includegraphics[width=0.4\textwidth]{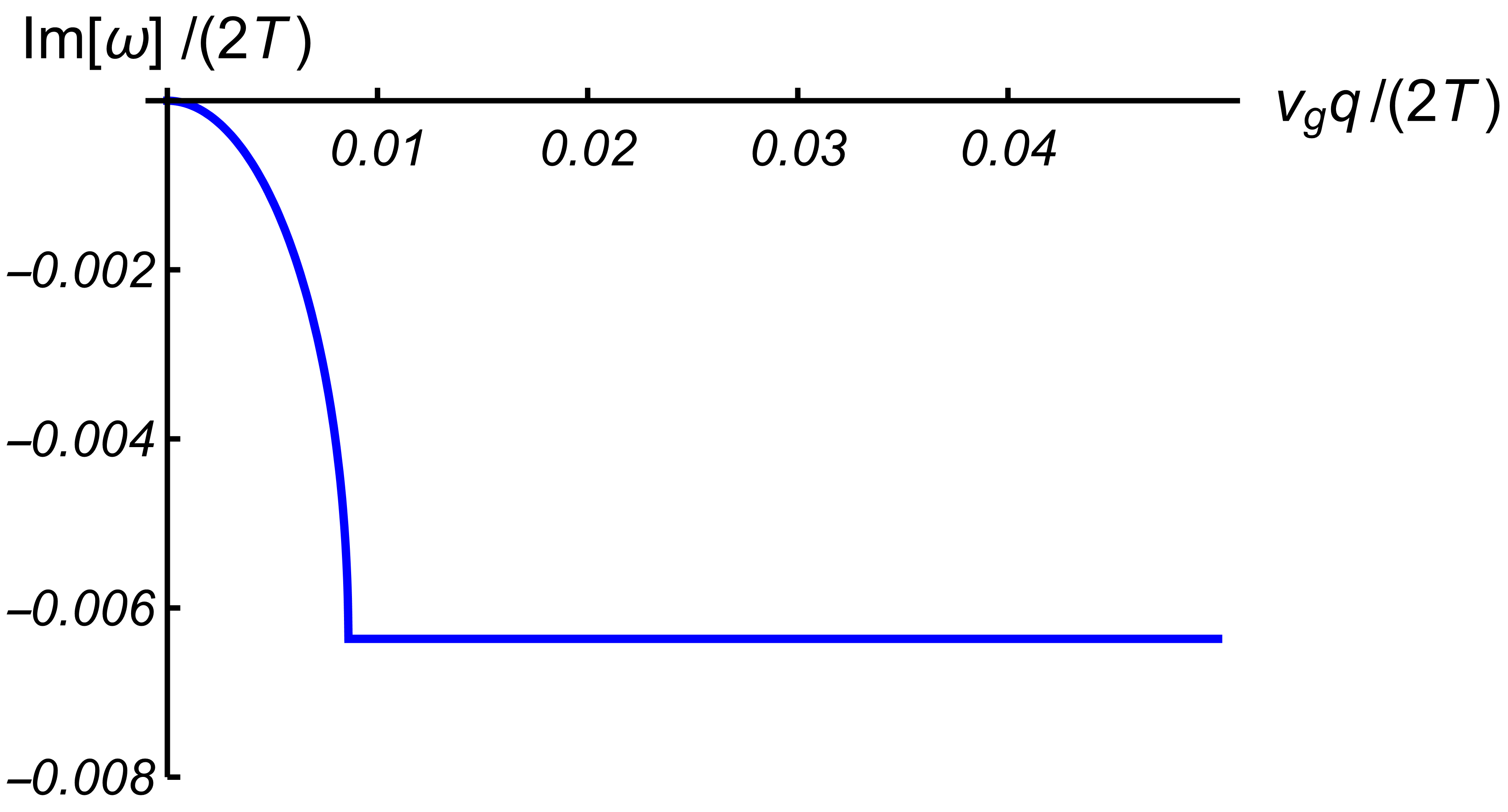}
}
\caption{Real and imaginary parts of the sound dispersion in strongly
  doped graphene in the presence of weak disorder, but neglecting
  viscosity. Top panels: the result for the Coulomb screening. Bottom
  panels: same for a gated structure. Dashed lines represents the
  ideal ``cosmic sound'' dispersion (\ref{cs0}). }
\label{fig9:sdcd}
\end{figure*}

\begin{figure*}[t]
\centerline{\includegraphics[width=0.3\textwidth]{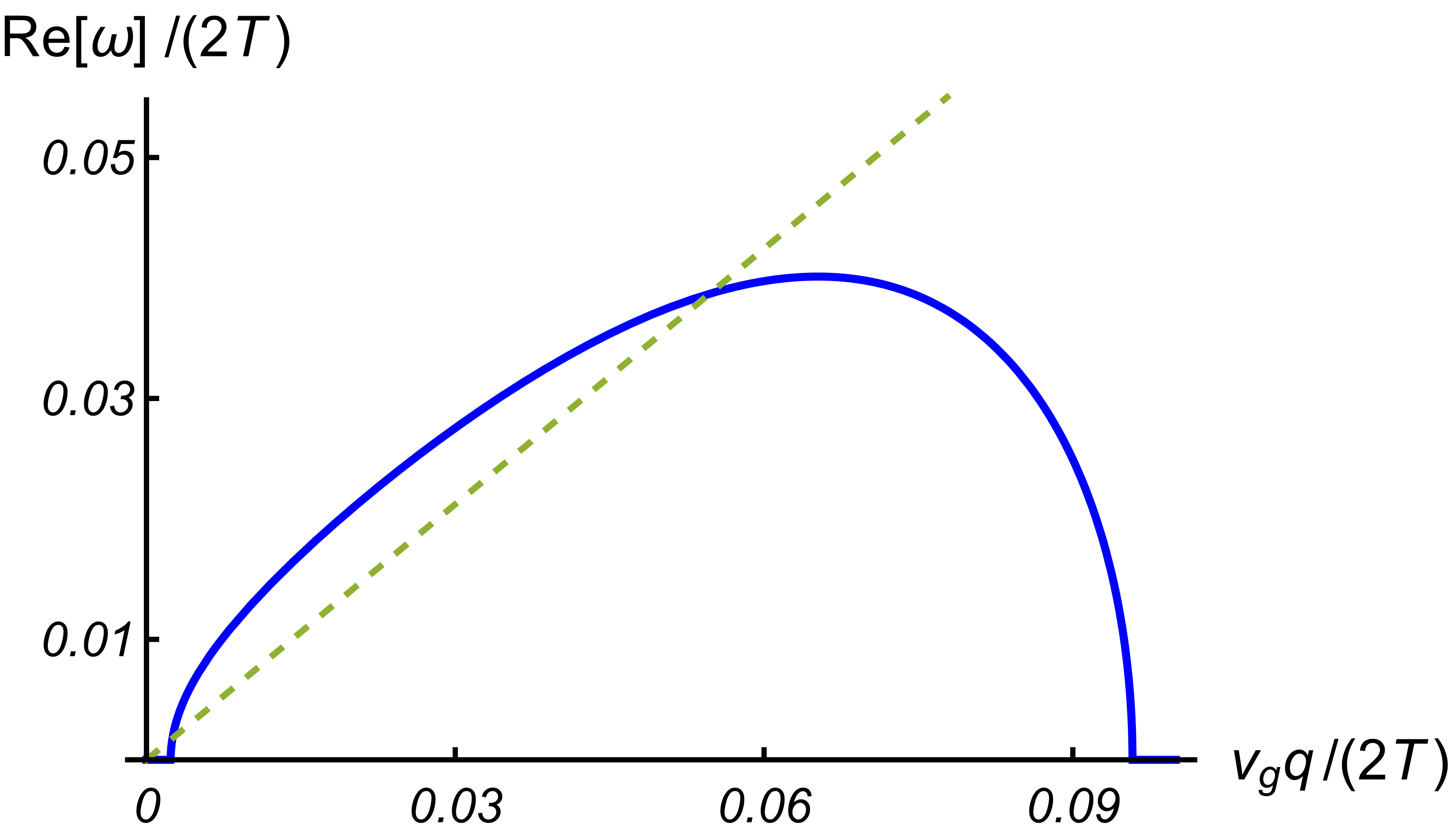}
\qquad
\includegraphics[width=0.3\textwidth]{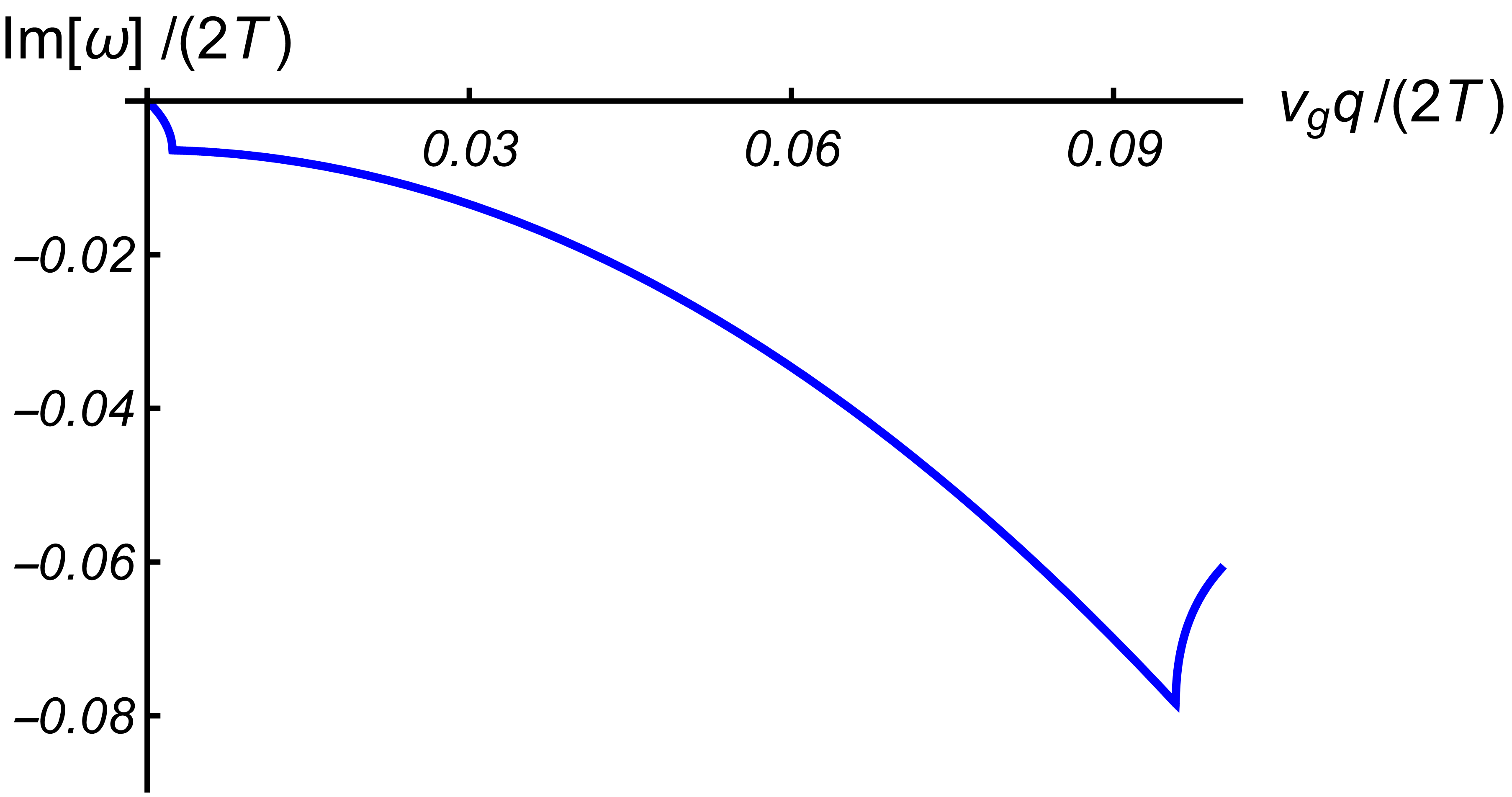}
\qquad
\includegraphics[width=0.3\textwidth]{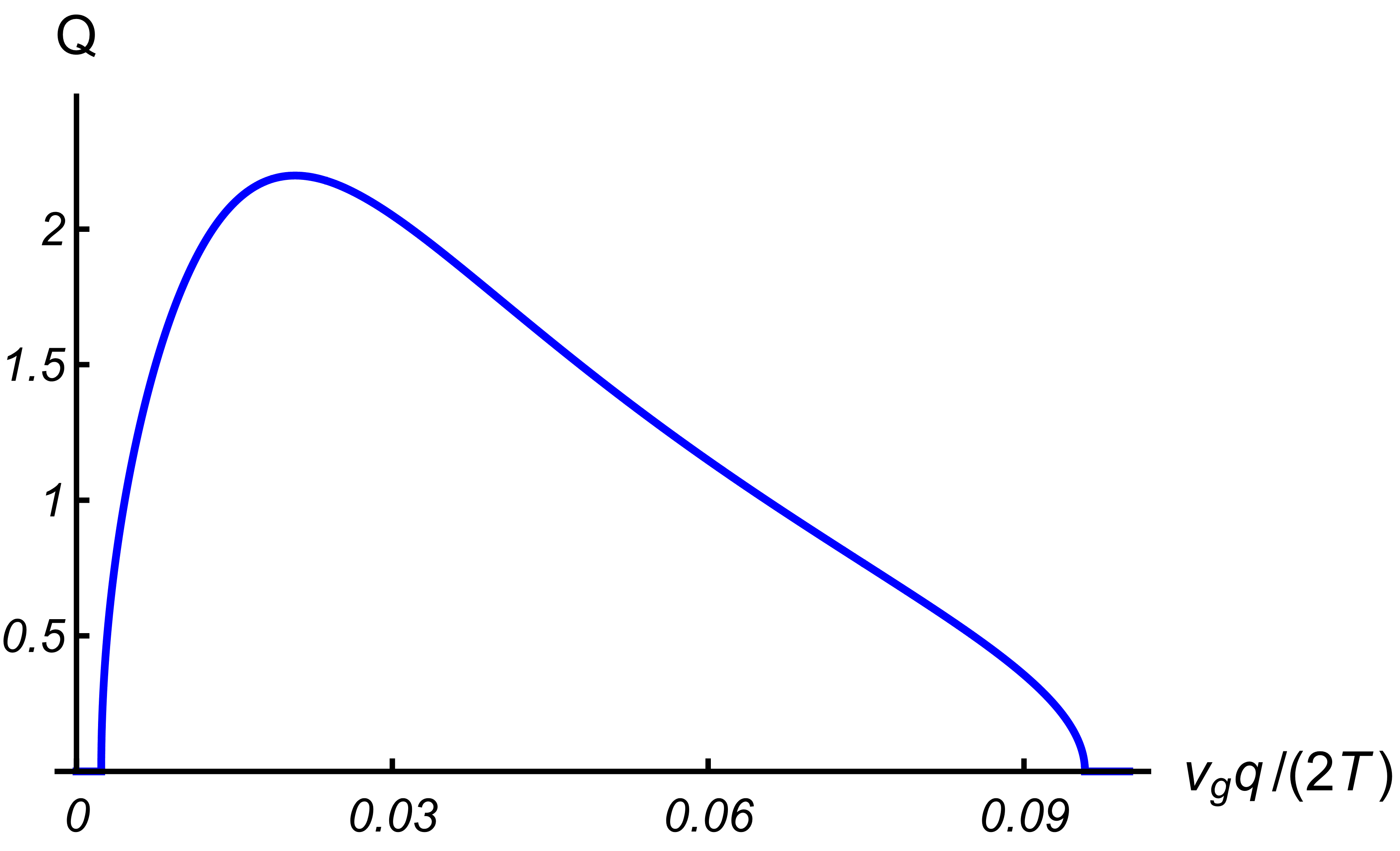}
}
\bigskip
\centerline{\includegraphics[width=0.3\textwidth]{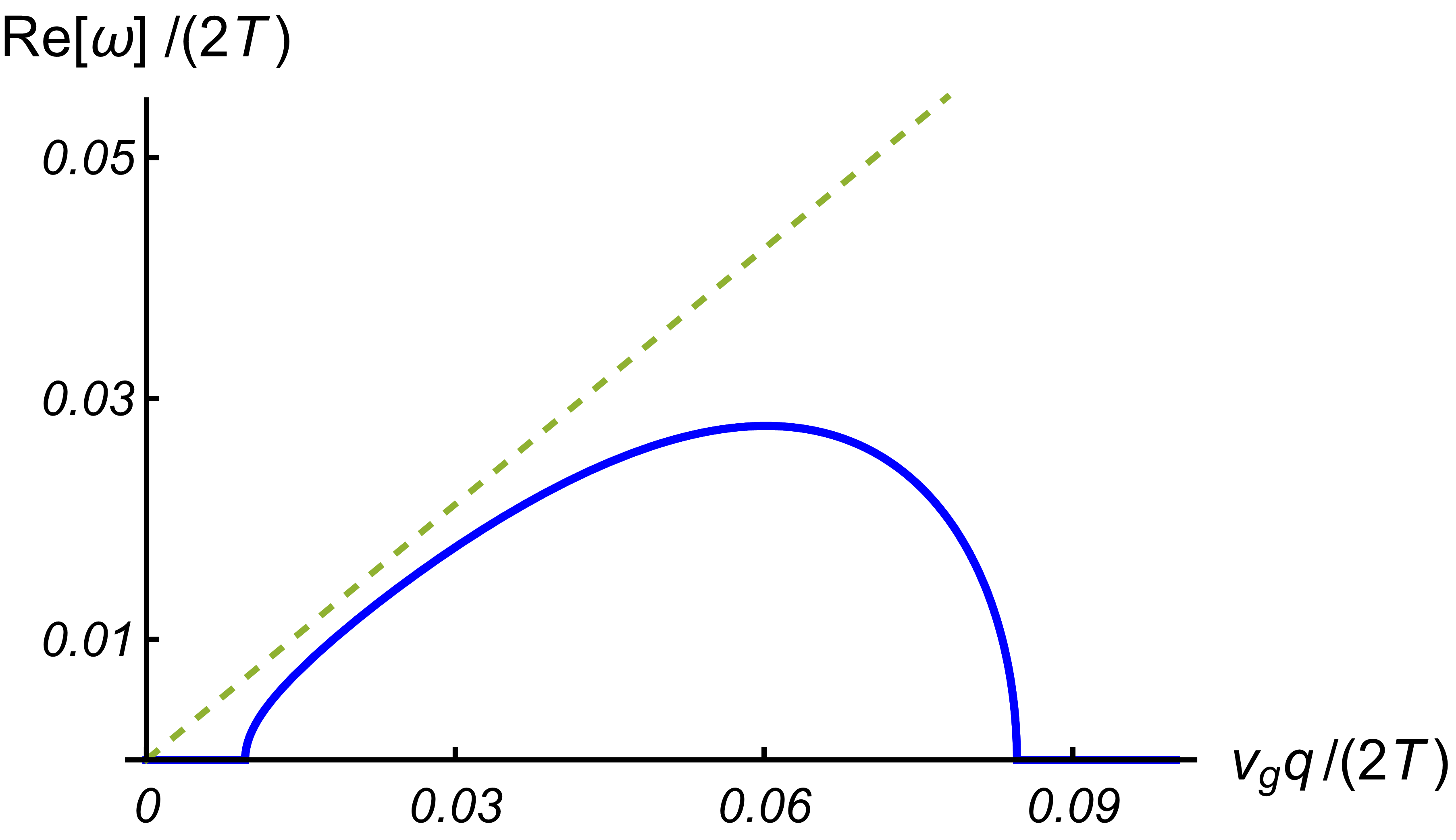}
\qquad
\includegraphics[width=0.3\textwidth]{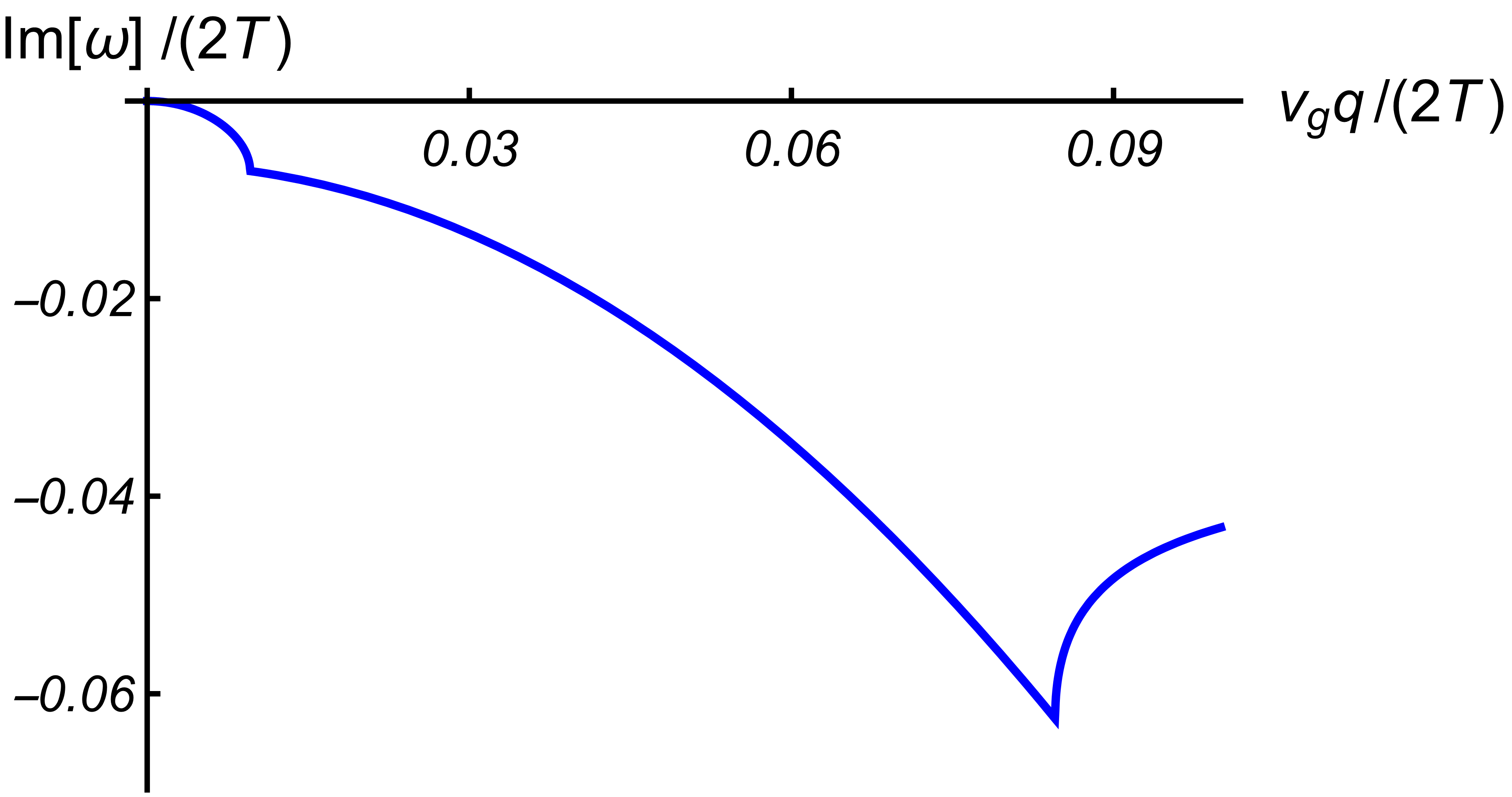}
\qquad
\includegraphics[width=0.3\textwidth]{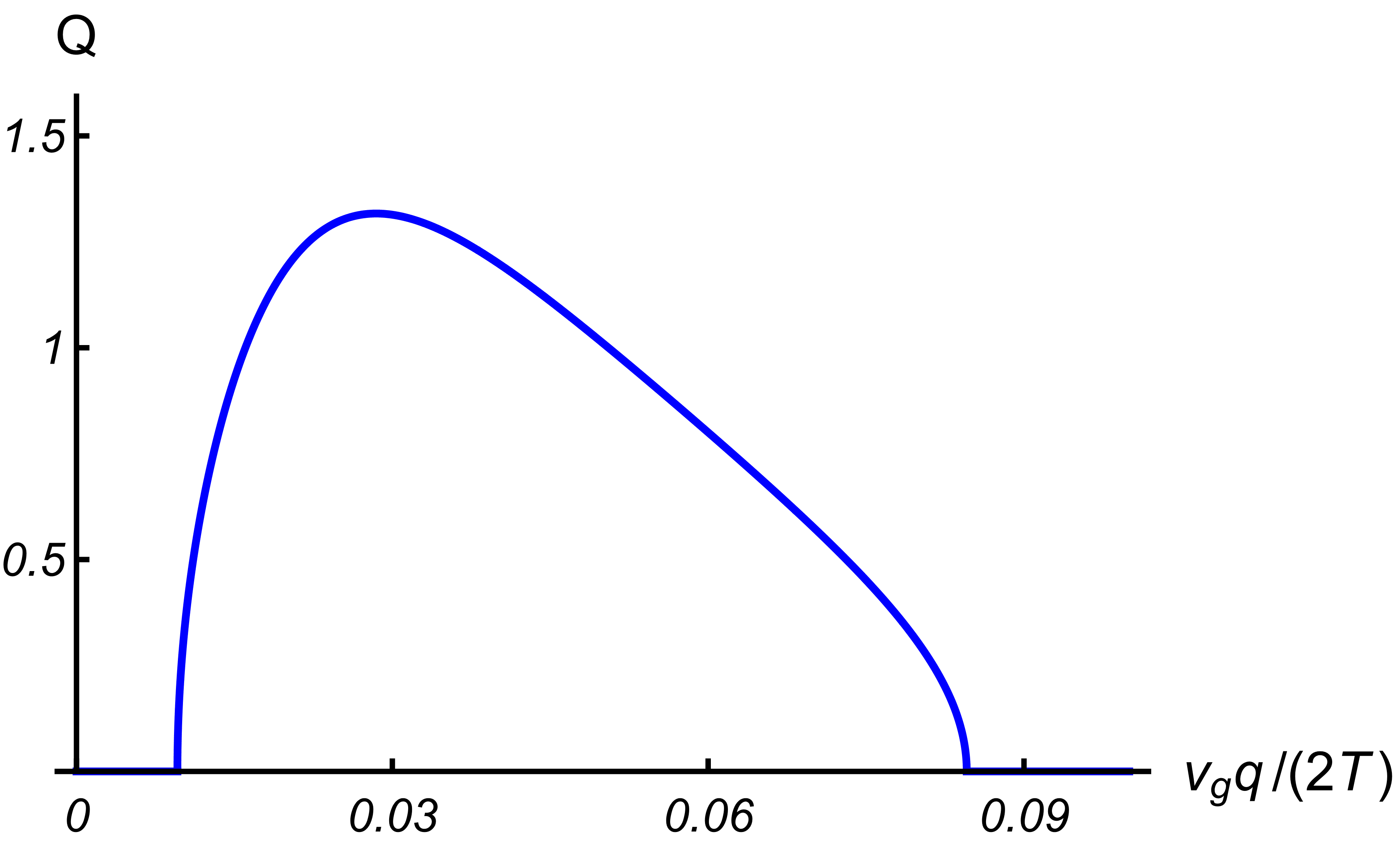}
}
\caption{Real and imaginary parts and the quality factor $Q={\rm
    Re}\,\omega/{\rm Im}\,\omega$ of the sound dispersion in strongly
  doped graphene in the presence of both weak disorder and
  viscosity. Top panels: the result for the Coulomb screening. Bottom
  panels: same for a gated structure. Dashed lines represent the ideal
  ``cosmic sound'' dispersion (\ref{cs0}). }
\label{fig10:sdcv}
\end{figure*}

Long-range Coulomb interaction modifies the screening contribution to
the sound mode (\ref{csflg})
\begin{equation}
\label{csflc}
\omega
\!=\!
\sqrt{\frac{v_g^2q^2}{2}
\!\left[1\!+\!\frac{\varkappa}{q}\right]
\!-\!\frac{\left(1
\!+\!\ell_G^2 q^2\right)^{\!2}}{4\tau^2_{{\rm dis}}}}
-
\frac{i\left(1
\!+\!\ell_G^2 q^2\right)^{\!2}}{2\tau_{{\rm dis}}}\!.
\end{equation}
Taking the naive limit $q\rightarrow0$ (and $x\rightarrow\infty$) in
Eq.~(\ref{csflc}), one arrives at the spectrum similar to the usual
two-dimensional plasmon \cite{hydro1,zna}
\begin{equation}
\label{cm30}
\omega(q\ll\varkappa) = - \frac{i}{2\tau_{{\rm dis}}}
+
\sqrt{\frac{1}{2}v_g^2q\varkappa-\frac{1}{4\tau^2_{{\rm dis}}}}.
\end{equation}
The dispersion (\ref{cm30}) is meaningful if the
following conditions are met
\[
q\ell_G\ll1, \quad q\ll\varkappa, \quad v_g^2\varkappa q \tau^2_{\rm dis} \gg 1.
\]
At the same time, for the hydrodynamic approach to be valid at all,
the gradients are supposed to be small on the scale that is defined by
the electron-electron interaction
\[
q\ell_{\rm hydro}\ll1, \quad
\ell_{\rm hydro}\sim\frac{v_g}{\alpha_g^2\bar{T}}.
\]
These conditions to be consistent if (using the explicit form of
physical quantities in the degenerate regime)
\[
v_g\varkappa\tau_{\rm dis}\gg1
\quad\Rightarrow\quad
N\alpha_g\mu\tau_{\rm dis}\gg1,
\]
\[
\ell_G\ll v_g^2\varkappa \tau^2_{\rm dis}
\quad\Rightarrow\quad
N^2\alpha^4_g\mu\tau_{\rm dis}(\bar{T}\tau_{\rm dis})^2\gg1,
\]
providing a possibility to observe the dispersion (\ref{cm30}) in a
parametrically defined range of wavevectors.

The eigenvectors of the ``flat zero mode'' and the sound mode mix the
charge, energy density, and velocity fluctuations. In that sense, the
mode (\ref{cm30}) is not a true plasmon, even though its dispersion is
identical with that of the usual plasmon in two dimensions. Moreover,
the dispersion (\ref{cm30}) resembles the plasmon dispersion only in
an intermediate interval of rather small $q$, while the true plasmon
exists at large values of $q$.

The above dispersion can be illustrated numerically as follows. Using
the same typical values $\tau_{\rm dis}^{-1}=1\,$THz,
$\nu=0.2\,$m$^2$/s (the kinematic viscosity varies only weakly with
the carrier density \cite{me2}), and $T=300\,$K, as well as the
typical value of the coupling constant \cite{gal,sav} $\alpha_g=0.23$
and the parameters characterizing the external gate in a typical
graphene-on-boron nitride structure \cite{geim1}, the dielectric
constant of the hexagonal boron nitride $\epsilon=4.4$ and the
graphene to gate distance $d=80\,$nm, we plot the two dispersions
(\ref{csflg}) and (\ref{csflc}) in
Figs.~\ref{fig8:sdc0}-\ref{fig10:sdcv}. In Fig.~\ref{fig8:sdc0}, we
show the two dispersions (\ref{csflg}) and (\ref{csflc}) in the
absence of both weak disorder and viscosity. The effect of the
screening can be summarized as follows. In a gated structure screening
leads to a slight (for the realistic parameter values chosen above)
change of slope of the sound mode dispersion. In contrast, Coulomb
screening leads to a plasmon-like square-root dispersion for the
smallest values of momentum, which soon turns into a linear dispersion
with the same slope as the ``cosmic sound'' of the ideal fluid, but
slightly (again, for the realistic parameter values) shifted upwards.
Taking into account dissipative processes washes out qualitative
differences between different types of screening. The results are also
qualitatively the same for strongly doped and neutral graphene. In
Fig.~\ref{fig9:sdcd} we show the results for the dispersion in the
presence of weak disorder, but still neglecting viscosity.
Qualitatively, the results for both types of screening are similar
with the only difference being that the real part of the dispersion in
the case of the Coulomb screening is shifted upwards relative to the
ideal sound dispersion, similarly to the left panel in
Fig.~\ref{fig8:sdc0}, while in the case of the gated structure the
resulting straight line at large enough $q$ has a slightly larger
slope than $1/\sqrt{2}$.

Once viscosity is taken into account, the curves in
Fig.~\ref{fig10:sdcv} strongly resemble the results in neutral
graphene, cf. Fig.~\ref{fig1:sd0v}. The results for gated graphene
show only insignificant numerical differences from the curves in
Fig.~\ref{fig1:sd0v}, while in the case of the Coulomb screening the
real part of the dispersion appears at a smaller value of $q$ and
exceeds the ideal spectrum (represented in all figures by the dotted
line) in a small intermediate range of $q$.

\section{Hydrodynamic collective modes and plasmons}
\label{lrt}

The hydrodynamic approach is applicable in the long-time and
long-wavelength limit \cite{dau10,luc,rev,me3}, i.e., at momenta that
are small compared to the typical ``equilibration'' length scale
$\ell_{\rm hydro}$. At higher momenta (and frequencies), the system is
not in equilibrium. In this regime (sometimes referred to
\cite{Giuliani} as ``collisionless''), the electronic fluid exhibits
well-known collective excitations, the plasmons. In two dimensions and
in the absence of impurity scattering
($\tau_{\rm{dis}}\rightarrow\infty$) the plasmon dispersion in the
degenerate electron gas has the form \cite{Giuliani}
\begin{equation}
\label{2Dplasmon}
\omega = \sqrt{2e^2\mu q}\left(1+\gamma \frac{q}{\varkappa}\right),
\end{equation}
where $\gamma$ is a numerical coefficient (see below). The ``proper''
way to derive Eq.~(\ref{2Dplasmon}) is to evaluate the Lindhard
function within the random phase approximation (RPA), which would lead
\cite{Giuliani} to the coefficient ${\gamma=3/4}$. An attempt to
derive the plasmon dispersion from a macroscopic (hydrodynamic-like)
theory leads to the same form (\ref{2Dplasmon}), but with a different
value for $\gamma$. This discrepancy is well known and can be
attributed to the failure of the hydrodynamic description at high
frequencies and momenta \cite{Giuliani}. As a result, one concludes
that the hydrodynamic collective modes have nothing to do with
plasmons simply because they belong to a different parameter
regime. In this Section we extend these arguments to Dirac fermions in
graphene and establish the relation between the above hydrodynamic
modes and plasmons.

\subsection{Degenerate regime}

The case of graphene is special because of the kinematic peculiarity
known as the ``collinear scattering singularity''
\cite{rev,luc,hydro1,hydro0,mfss,mfs,schutt,drag2,me1} leading to the
existence of the two parametrically (in the weak coupling limit)
different length scales associated with electron-electron
interactions, ${\ell_{\rm coll}\ll\ell_{\rm hydro}}$. In an
intermediate momentum range,
${\ell_{\rm{hydro}}^{-1}\ll{q}\ll\ell_{\rm coll}^{-1}}$, the
hydrodynamic theory of Section~\ref{ht} breaks down, while the linear
response theory of Ref.~\onlinecite{hydro0} is still
valid. Remarkably, the macroscopic equations of the latter theory are
identical with the linearized hydrodynamic equations, so that the
collective modes in the two parameter regimes coincide.

In the degenerate regime and in the absence of magnetic field, the
linear response theory \cite{hydro0} reduces to the single macroscopic
equation describing the dynamics of the electric current $\bs{J}$
(here $\rho$ is the charge density)
\begin{equation}
\label{ceqlr}
\frac{\partial\bs{J}}{\partial t} + \frac{v_g^2}{2}\bs{\nabla}\rho
-\nu \Delta\bs{J}
-\frac{v_g^2}{2}\frac{\partial n}{\partial\mu} e^2 \bs{E}
= - \frac{\bs{J}}{\tau_{\rm dis}},
\end{equation}
which is essentially the generalized Ohm's law. To obtain the plasmon
dispersion, we introduce the Vlasov field [cf. Eq.~(\ref{vlasov})] and
use the continuity equation. In the case of Coulomb interaction, the
standard algebra \cite{Giuliani} leads to the following equation
\[
\omega\left(1+q^2\ell_G^2-i\omega\tau_{\rm dis}\right) 
=
-iDq^2 - i2\pi \sigma q,
\]
where ${D=v_g^2\tau_{\rm dis}/2}$ and ${\sigma=v_g^2(\partial
  n/\partial\mu)\tau_{\rm dis}/2}$ are the diffusion coefficient and
the Drude conductivity. The resulting spectrum has the form
\begin{equation}
\label{flcm}
\omega = 
\sqrt{2e^2\mu q\left(1\!+\!\frac{q}{\varkappa}\right)-\frac{(1\!+\!q^2\ell_G^2)^2}{4\tau^2_{\rm dis}}}
-\frac{i(1\!+\!q^2\ell_G^2)}{2\tau_{\rm dis}}.
\end{equation}
The spectrum (\ref{flcm}) is exactly the same as
Eq.~(\ref{csflc}). For a clean system
(${\tau_{\rm{dis}}\rightarrow\infty}$), the expansion for small
${q\rightarrow0}$ yields the form (\ref{2Dplasmon}) with the ``wrong''
coefficient, ${\gamma=1/2}$. At the same time, the leading term
(neglecting the correction for ${q\ll\varkappa}$) agrees with the
standard Fermi liquid result even in the presence of disorder
\cite{zna} (neglecting viscosity).

The expression (\ref{flcm}) is valid for momenta up to $\ell_{\rm coll}^{-1}$,
but in fact it becomes overdamped already at momenta of order
$\ell_{\rm hydro}^{-1}$. At larger momenta, $q\gg\ell_{\rm coll}^{-1}$, the
quasi-equilibrium description breaks down and the true plasmons emerge
with the dispersion (\ref{2Dplasmon}). By that time the spectrum
(\ref{flcm}) becomes purely imaginary (see Fig.~\ref{fig10:sdcv}), and
hence the two modes are not connected. Similar conclusions have been
reached in Ref.~\onlinecite{ldsp}, where it was argued that Coulomb
interaction precludes the appearance of hydrodynamic sound in Fermi
liquids.

\subsection{Two-fluid hydrodynamics}
\label{tfm}

Let us slightly digress and consider the curious case of the two-fluid
hydrodynamics \cite{mr3,mr2,cfl} in compensated semimetals. Following
Ref.~\onlinecite{mr2} we assume that the full electronic systems
comprises two weakly coupled fluids, one consisting of electrons and
another of holes. This means that the length scales $\ell_{ee}$ and
$\ell_{hh}$ describing intraband electron-electron scattering are much
smaller than the interband scattering length $\ell_{eh}$. In that
case, the system is described by two equations similar to
Eq.~(\ref{ceqlr}) with an extra interband scattering term
\begin{eqnarray}
\label{ceq2f}
&&
\frac{\partial\bs{j}_\alpha}{\partial t} + \frac{\langle v^2\rangle}{2}\bs{\nabla}n_\alpha
-\nu \Delta\bs{j}_\alpha
-\frac{\langle v^2\rangle}{2}\frac{\partial n_\alpha}{\partial\mu} e_\alpha \bs{E} =
\\
&&
\nonumber\\
&&
\qquad\qquad\qquad\qquad\qquad\qquad\quad
= - \frac{\bs{j}_\alpha}{\tau_{\rm dis}} - \frac{\bs{j}_\alpha-\bs{j}_{\alpha'}}{2\tau_{eh}},
\nonumber
\end{eqnarray}
where $e_h=-e>0$, $e_e=e<0$, $\bs{j}_\alpha$ denotes the quasiparticle
currents, and $\alpha'$ denotes the other constituent. For simplicity
we assume the system to be electron-hole symmetric
($\ell_{ee}=\ell_{hh}$).

Combining the two currents into the linear combinations,
${\bs{j}=\bs{j}_e-\bs{j}_h}$ and ${\bs{j}_I=\bs{j}_e+\bs{j}_h}$, we
find the decoupled (in the absence of the magnetic field) equations
\begin{subequations}
\label{ceqs2fpj}
\begin{equation}
\label{ceq2fj}
\frac{\partial\bs{j}}{\partial t} 
-\nu \Delta\bs{j}
-\frac{\langle v^2\rangle}{2}\frac{\partial n_I}{\partial\mu} e \bs{E} 
= - \frac{\bs{j}}{\tau_{\rm dis}} - \frac{\bs{j}}{\tau_{eh}},
\end{equation}
\begin{equation}
\label{ceq2fp}
\frac{\partial\bs{j}_I}{\partial t} + \frac{\langle v^2\rangle}{2}\bs{\nabla}n_I
-\nu \Delta\bs{j}_I
= - \frac{\bs{j}_I}{\tau_{\rm dis}}.
\end{equation}
\end{subequations}
Combining these equations with the two continuity equations
(\ref{cen1}) and (\ref{ceni1}), we find a sound-like mode 
\begin{subequations}
\label{2fms}
\begin{equation}
\label{2fs}
\omega = 
\sqrt{\frac{\langle v^2\rangle q^2}{2} 
\!-\!\left(\frac{1\!+\!q^2\ell_G^2}{2\tau_{\rm dis}}\!-\!\frac{1}{2\tau_R}\right)^{\!\!2}}
-\frac{i(1\!+\!q^2\ell_G^2)}{2\tau_{\rm dis}}-\frac{i}{2\tau_R},
\end{equation}
and a plasmon-like mode
\begin{equation}
\label{2fp}
\omega = 
\sqrt{\frac{\langle v^2\rangle \varkappa q}{2}
-\frac{(1\!+\!q^2\ell_{G*}^2)^2}{4\tau^2_{*}}}
-\frac{i(1\!+\!q^2\ell_{G*}^2)}{2\tau_{*}},
\end{equation}
where
\[
\varkappa = 2\pi e^2 \frac{\partial n_I}{\partial\mu},
\qquad
\tau_* = \frac{\tau_{\rm dis}\tau_{eh}}{\tau_{\rm dis}\!+\!\tau_{eh}},
\qquad
\ell_{G*} = \sqrt{\nu\tau_*}.
\]
\end{subequations}

In the hydrodynamic parameter range, both modes (\ref{2fms}) are well
defined. The expression under the square root in Eq.~(\ref{2fs}) can
be rewritten as
\[
\frac{\langle v^2\rangle}{2} 
\left[
q^2
\left(1\!-\!\frac{\tau_{ee}}{\tau_{\rm dis}}\!+\!\frac{\tau_{ee}}{\tau_R}
\!-\!\frac{q^2\ell_{ee}^2}{2}\right)
\!-\!
\frac{1}{2}\left(\frac{1}{\ell_{\rm dis}}\!-\!\frac{1}{\ell_R}\right)^2\right]\!.
\]
Here ${\tau_{ee}\ll\tau_{\rm dis}}$, ${\tau_{ee}\ll\tau_R}$ by the
assumptions of the hydrodynamic regime and ${q\ell_{ee}\ll1}$ under
the assumption of the gradient expansion in the hydrodynamic theory
(here we consider a generic semimetal and hence do not have the
aforementioned scale separation specific to graphene, hence we cannot
extend the argument beyond the validity region of the gradient
expansion). Therefore apart from the small gap due to the interplay
between disorder scattering and recombination processes, the sound
mode is well defined within the hydrodynamic range of momenta.

Similar arguments can be extended to the plasmon-like mode
(\ref{2fp}). Assuming a clean system, ${\tau_{eh}\ll\tau_{\rm dis}}$,
${\tau_*\rightarrow\tau_{eh}}$, one finds under the square root in
Eq.~(\ref{2fp})
\[
\frac{\langle v^2\rangle}{2} 
\left[
q\varkappa
\!-\!q^2\frac{\tau_{ee}}{\tau_{eh}}
\!-\!\frac{q^4\ell_{ee}^2}{2}
\!-\!
\frac{1}{2\ell_{eh}^2}\right]\!.
\]
Typically, the Thomas-Fermi screening radius is smaller then the
electron-electron scattering length,
${\varkappa\ell_{ee}\gg1}$. Hence, the mode (\ref{2fp}) is also well
defined. Here the electron-hole scattering yields the (small) gap in
the dispersion similarly to the disorder scattering in
Eq.~(\ref{flcm}).

\subsection{Graphene at charge neutrality}

Utilizing the scale separation in graphene (see above), we can
approach the question of the collective modes from the standpoint of
the linear response theory of Ref.~\onlinecite{hydro0}. Here, instead
of formulating the hydrodynamic equations (\ref{hydro}), we turn to
the macroscopic equations describing the behavior of the three
inequivalent currents in the system, $\bs{j}$, $\bs{j}_I$, and
$\bs{j}_E$
\begin{subequations}
\label{lreqsdp}
\begin{equation}
\label{ceqlrdp}
\frac{\partial\bs{j}}{\partial t} + \frac{v_g^2}{2}\bs{\nabla}n
-\frac{2\ln2}{\pi}e^2T \bs{E}
= - \frac{\bs{j}}{\tau_{\rm dis}} - \frac{\bs{j}}{\tau_{11}},
\end{equation}
\begin{equation}
\label{iceqlrdp}
\frac{\partial\bs{j}_I}{\partial t} + \frac{v_g^2}{2}\bs{\nabla}n_I
-\frac{\gamma_1\nu}{T} \Delta\bs{j}_E
= - \frac{\bs{j}_I}{\tau_{\rm dis}} - \frac{\bs{j}_I-\frac{\pi^2\bs{j}_E}{27\zeta(3)T}}{\tau_{22}\delta_I},
\end{equation}
\begin{equation}
\label{eceqlrdp}
\frac{\partial\bs{j}_E}{\partial t} + \frac{v_g^2}{2}\bs{\nabla}n_E
-\nu \Delta\bs{j}_E
= - \frac{\bs{j}_E}{\tau_{\rm dis}},
\end{equation}
\end{subequations}
where $\gamma_1$ is a numerical prefactor. At charge neutrality, the
viscous term vanishes from Eq.~(\ref{ceqlrdp}) in contrast to the
two-fluid model, see Eq.~(\ref{2fs}). In graphene, the electron and
hole subsystems are strongly coupled
(${\ell_{ee}=\ell_{hh}\sim\ell_{eh}}$) forming a single fluid, where
the electric current is not affected by viscous effects because of
electron-hole symmetry. Viscosity still affects neutral quasiparticle
and energy flows in agreement with the hydrodynamic approach, where
the hydrodynamic velocity in neutral graphene describes the flow of
energy.

\begin{figure*}[t]
\centerline{\includegraphics[width=0.46\textwidth]{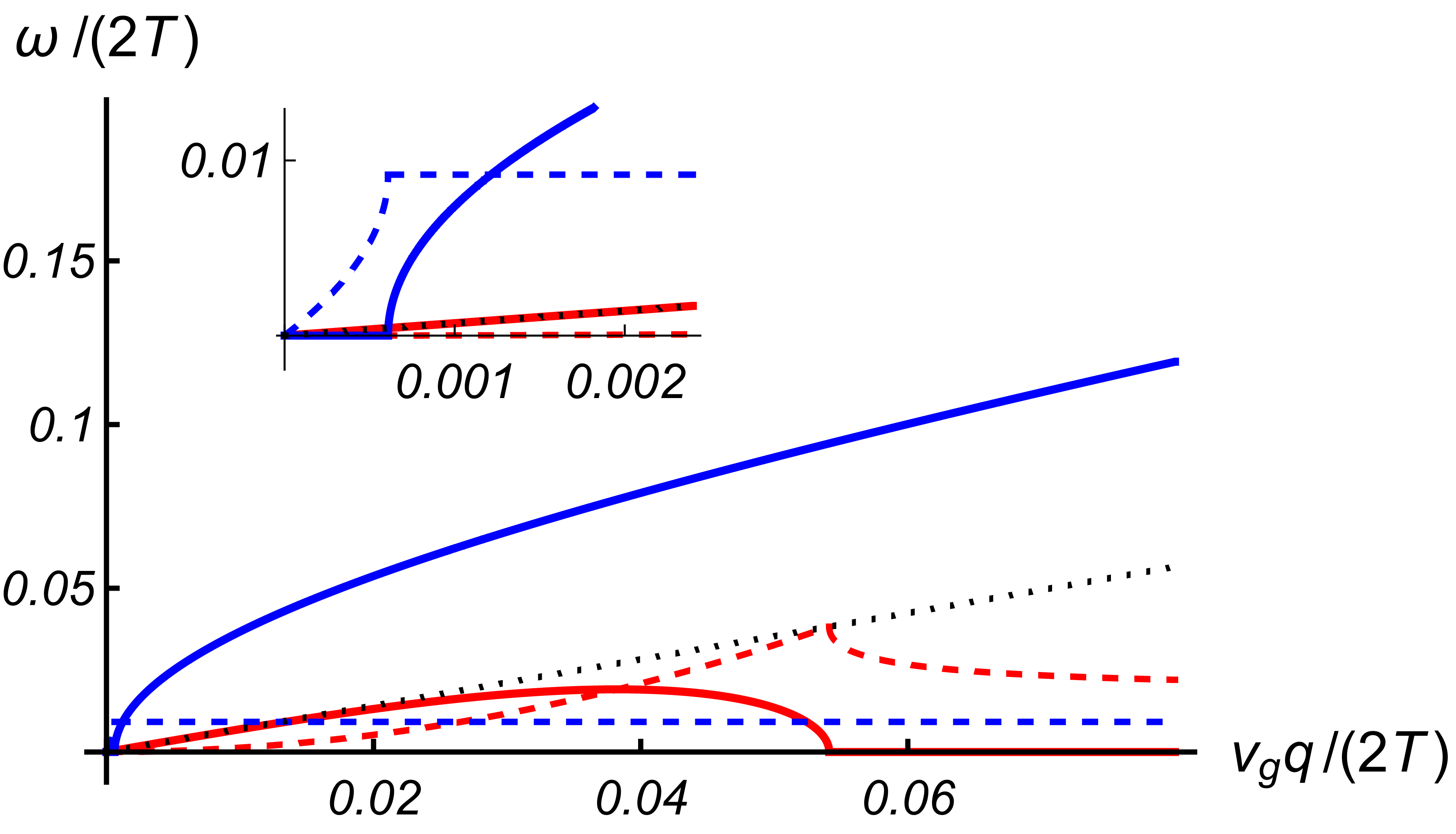}
\quad
\includegraphics[width=0.46\textwidth]{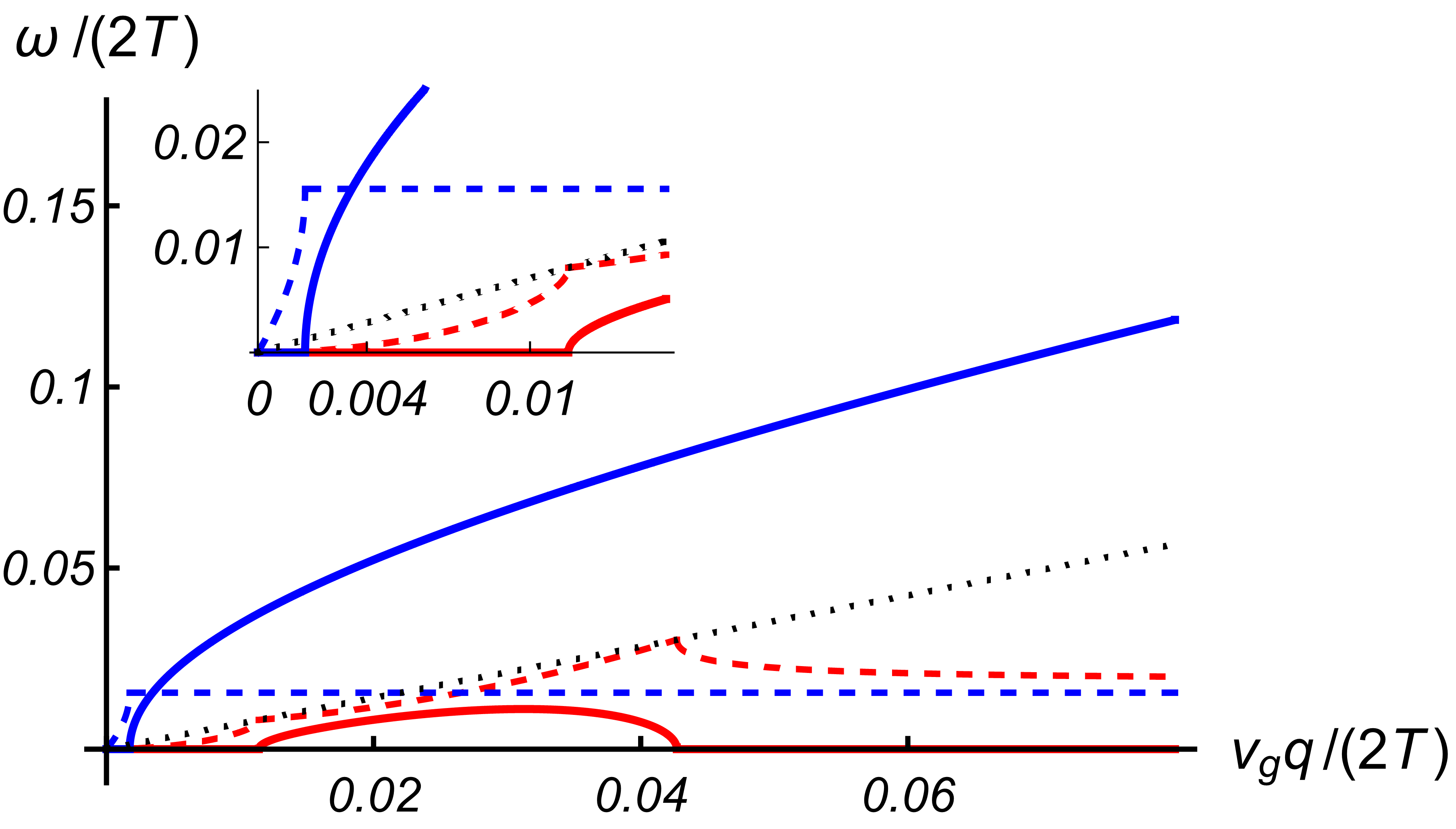}
}
\caption{Comparison between the plasmon mode (\ref{pdd}) and the sound
  mode (\ref{sdd}) within the linear response theory. Solid curves
  show the real part of the dispersion, dashed curves the absolute
  value of the imaginary part. The dotted line shows the ideal
  ``cosmic sound'' dispersion (\ref{cs0}). The plasmon dispersion is
  shown in blue, the sound in red. The distinction between the two
  modes is clearly defined by their frequencies that are much higher
  for the plasmon mode. Left panel shows the dispersion for a clean
  sample; right panel the same for the typical value $\tau_{\rm
    dis}^{-1}=1\,$THz. The coupling constant is taken at a model value
  $\alpha_g=0.1$, hence, no renormalization of the velocity $v_g$ is
  taken into account strongly underestimating viscosity. The real part
  of the sound dispersion vanishes at $\tilde{q}\approx0.54$, which is
  similar to the applicability limit of the linear response theory,
  $\ell_{\rm coll}^{-1}$. The imaginary part exceeds the real part at
  a lower value of $\tilde{q}$, such that the mode becomes overdamped
  and disappears still within the applicability region of the
  theory. In the presence of disorder (right panel) the sound model is
  completely overdamped, see Fig.~\ref{fig1:sd0v} for more realistic
  values.}
\label{fig11:pl}
\end{figure*}

Similarly to the hydrodynamic regime (Section~\ref{hdp}), the energy
and charge decouple completely. Combining Eq.~(\ref{eceqlrdp}) with
the continuity equation for the energy density (\ref{tteqlin1}) --
that is equivalent to the linearized heat transport equation
(\ref{eqent}) -- we recover the sound mode (\ref{csv}). 

On the other hand, combining Eq.~(\ref{ceqlrdp}) with the continuity
equation (\ref{cen1}) we find
\begin{equation}
\label{pldp0}
\omega^2 +i\omega\left(\frac{1}{\tau_{\rm dis}}+\frac{1}{\tau_{11}}\right)
=
\frac{v_g^2}{2}q^2 + (4\ln2)e^2Tq,
\end{equation}
leading to the plasmon-like spectrum. For large enough frequencies,
${\omega\gg\tau_{11}^{-1}\gg\tau_{\rm dis}^{-1}}$, and small momenta,
${q\rightarrow0}$, the resulting dispersion coincides with the leading
behavior of the true plasmon dispersion established in
Ref.~\onlinecite{schutt}
\begin{equation}
\label{pldp}
\omega = \sqrt{(4\ln2)e^2Tq}
\quad\Rightarrow\quad
\tilde\omega = \sqrt{2(\ln2)\alpha_g\tilde{q}},
\end{equation}
where the last equality is expressed in terms of the dimensionless
variables (\ref{dqs}), also used in Ref.~\onlinecite{schutt}. Note,
that at large momenta, where the first term in the left-hand side of
Eq.~(\ref{pldp0}) dominates, the resulting dispersion resembles the
cosmic sound (\ref{cs0}), contradicting the result of
Ref.~\onlinecite{schutt}, where the dispersion in the large-$q$ limit
also becomes linear, but without the extra $\sqrt{2}$.

Considering the limit $\tau_*\rightarrow\infty$ in Eq.~(\ref{2fp}), we
arrive at the same result [in graphene at the charge neutrality point,
  ${v_g^2\varkappa/2=(4\ln2)e^2T}$, while viscosity does not affect
  charge transport]. In the absence of disorder, the two-fluid model
considered in Section~\ref{tfm} describes the electron and hole
subsystems as being weakly coupled (similarly to the effect of Coulomb
drag \cite{dragrev}, but without spatial separation). Charge density
fluctuations are correspond to the out-of-phase motion of electrons
and holes. In the absence of the electron-hole scattering
($\tau_*\rightarrow\infty$), charge transport is effectively decoupled
from the in-phase (imbalance) mode and hence Eq.~(\ref{ceq2fj})
becomes equivalent to Eq.~(\ref{ceqlrdp}) yielding the same plasmonic
mode.

Rewriting Eq.~(\ref{pldp0}) in the form
\[
i\omega\left[-i\omega +\frac{1}{\tau_{\rm dis}}+\frac{1}{\tau_{11}}\right]
=
\frac{v_g^2}{2}q^2 + (4\ln2)e^2Tq,
\]
we express the plasmon dispersion in the form closely resembling 
Eq.~(\ref{cmdp})
\[
\omega = -i \frac{\sigma(\omega)q^2}{e^2\partial n/\partial\mu}
\left[1 + e V_s(q)\frac{\partial n}{\partial\mu}\right],
\]
where instead of the static conductivity (\ref{sq}) we find the optical
conductivity \cite{me3}
\[
\sigma(\omega) = \frac{2e^2T\ln2}{\pi} 
\frac{1}{-i\omega + \tau_{11}^{-1}+\tau_{\rm dis}^{-1}}.
\]
In the hydrodynamic regime
${\sigma(\omega\rightarrow0)\rightarrow\sigma_0}$ and we recover the
diffusive mode (\ref{cmdp}).

Resolving Eq.~(\ref{pldp0}) we find the full plasmon dispersion
\begin{equation}
\label{pd}
\omega = -i\frac{\tau_{\rm dis}\!+\!\tau_{11}}{2\tau_{\rm dis}\tau_{11}}
\!+\!
\sqrt{\frac{v_g^2}{2}q^2 \!+\! (4\ln2)e^2Tq 
\!-\! \frac{(\tau_{\rm dis}\!+\!\tau_{11})^2}{4\tau_{\rm dis}^2\tau_{11}^2} }\ .
\end{equation}
To analyze the two modes -- the plasmon and sound -- together, we
rewrite the above dispersion in dimensionless units (\ref{dqs}). The
plasmon dispersion takes the form
\begin{eqnarray}
&&
\tilde\omega_p\!=\!\sqrt{2(\ln2)\alpha_g\tilde{q}\left[1\!+\!\frac{\tilde{q}}{4(\ln2)\alpha_g}\right]\!-\!
\left[\frac{1}{2\tilde\tau_{\rm dis}}\!+\!\frac{\alpha_g^2\ln2}{2\pi{\cal A}}\right]^2}
\nonumber\\
&&
\nonumber\\
&&
\qquad\qquad\qquad\qquad
-
\frac{i}{2}\!\left(\frac{1}{\tilde\tau_{\rm dis}}\!+\!\frac{\alpha_g^2\ln2}{\pi{\cal A}}\right)\!,
\label{pdd}
\end{eqnarray}
where the constant ${{\cal A}\approx0.12}$ determines the quantum
conductivity at charge neutrality \cite{kash,me1,rev,luc}
\[
\sigma_Q={\cal A}e^2/\alpha_g^2.
\]
At the same time, the sound dispersion (\ref{cs0}) is given by
\begin{equation}
\label{sdd}
\tilde\omega_s\!=\!
\sqrt{\frac{\tilde{q}^2}{2}\!-\!
\left[\frac{1}{2\tilde\tau_{\rm dis}}\!+\!\frac{\pi{\cal B}\tilde{q}^2}{9\zeta(3)\alpha_g^2}\right]^2}
-
\frac{i}{2}\!\left(\frac{1}{\tilde\tau_{\rm dis}}\!+\!\frac{2\pi{\cal B}\tilde{q}^2}{9\zeta(3)\alpha_g^2}\right)\!,
\end{equation}
where the constant ${{\cal B}\approx0.45}$ determines the shear
viscosity in neutral graphene \cite{msf,me1,rev,luc}
\[
\eta(\mu\!=\!0) = {\cal B}T^2/(\alpha_g^2v_g^2).
\]

In pure graphene (${\tilde\tau_{\rm dis}\rightarrow\infty}$) in the
weak coupling limit (${\alpha_g\rightarrow0}$), the regions where the
two dispersions are real overlap: the plasmon dispersion (\ref{pdd})
is real for $\tilde{q}\gg\alpha_g^3$, while the sound dispersion
(\ref{sdd}) is real for $\tilde{q}\ll\alpha_g^2$. Weak disorder does
not yield any qualitative changes.

The linear response theory, Eqs.~(\ref{lreqsdp}), is applicable at
length scales larger than $\ell_{\rm coll}$, the graphene-specific
scale [see Eq.~(\ref{scasep})], reflecting the collinear scattering
singularity. In dimensionless units,
${\ell_{\rm{coll}}^{-1}\sim\alpha_g^{2}|\ln\alpha_g|}$, which in the
weak coupling limit greatly exceeds
${\ell_{\rm{hydro}}^{-1}\sim\alpha_g^{2}}$, which determines the
applicability of the hydrodynamic theory of Section~\ref{ht}. In the
limit ${\tilde\tau_{\rm dis}\rightarrow\infty}$, the real part of the
sound dispersion (\ref{sdd}) vanishes when
\[
\tilde{q} = \tilde{q}_0 = \frac{9\zeta(3)}{\sqrt{2}\pi{\cal B}} \alpha_g^2
\approx 5.41 \alpha_g^2.
\]
Here the large numerical coefficient may mask the difference between
the two length scales $\ell_{\rm hydro}$ and $\ell_{\rm coll}$ for all
but the lowest values of $\alpha_g$. We illustrate the resulting
dispersions in Fig.~\ref{fig11:pl}, where we use a model value
$\alpha_g=0.1$ to keep the two length scales well separated. Even
though $\tilde{q}_0$ is of the same order of magnitude as $\ell_{\rm
  coll}$, the imaginary part of the dispersion becomes comparable to
the real part at a significantly lower value of $\tilde{q}$. At that
point the mode becomes overdamped and essentially disappears. Adding
realistic disorder renders the mode completely overdamped, see the
right panel in Fig.~\ref{fig11:pl}.

\section{Summary}

In this paper we described electronic collective modes in graphene
based on the hydrodynamic approach and compared the results with the
more general linear response theory. Our results generalize the
discussion of these issues reported in Ref.~\onlinecite{hydro1} within
the small momentum expansion. Given the universality of hydrodynamics,
the results for the collective modes in the hydrodynamic regime are
applicable to other semimetals (where the momentum density represented
by $\bs{u}$ is effectively decoupled from the charge transport unless
the system is doped far away from charge neutrality), while the
three-mode approximation used to derive the linear response theory
discussed in Section~\ref{lrt} is specific to graphene.

Our main results are illustrated in Figs.~\ref{fig1:sd0v} and
\ref{fig11:pl}. The former shows the dispersion of the sound mode in
the hydrodynamic regime with the viscous damping and weak disorder
taken into account. Using the typical experimental values of the
viscosity and disorder scattering time, we find that the sound mode in
real graphene is strongly damped, making it difficult to observe the
ideal ``cosmic sound'' dispersion (\ref{cs0}) experimentally.

In Fig.~\ref{fig11:pl} we illustrate the sound and plasmon modes in
neutral graphene obtained within the linear response theory of
Ref.~\onlinecite{hydro0} (extended beyond the stationary and uniform
fields). Both modes are evaluated with the ``bare'' parameter values
(ignoring, e.g., the renormalization of quasiparticle spectrum in
graphene \cite{me2,shsch}) for clarity. Effectively, this approach
strongly underestimates the kinematic viscosity and hence the sound
mode in Fig.~\ref{fig11:pl} is much more pronounced than in
Fig.~\ref{fig1:sd0v}.

The plasmon mode (\ref{pdd}) is characterized by higher frequencies
that the sound mode (\ref{sdd}) and hence is not accessible within the
standard hydrodynamic approach of Section~\ref{ht}. The connection
between the two calculations can be made by allowing for the
frequency-dependent (optical) conductivity in Eqs.~(\ref{detfull}) and
(\ref{cmdp}). Reducing the dissipative coefficients in the
hydrodynamic theory to frequency-independent constants (following the
standard approach of Ref.~\onlinecite{dau6}) leads to the diffusive
behavior of the collective charge fluctuations, see
Eq.~(\ref{cmdp}). Similarly, all other hydrodynamic collective modes
(except for the sound mode) are characterized by purely imaginary
spectra. This should be contrasted with the linear response theory,
Eqs.~(\ref{lreqsdp}), that allows for the frequency-dependent
conductivities leading to the real plasmon dispersion (\ref{pdd}), as
well as a third (neutral) collective mode following from
Eqs.~(\ref{iceqlrdp}) and (\ref{ceni1}). The fact that these
additional (to the sound) modes can be reached within the linear
response theory and connected to the hydrodynamic description should
be attributed to the scale separation in graphene (due to the
kinematic peculiarity of Dirac fermions \cite{mfs,mfss,rev,luc,me1}),
see Eq.~(\ref{scasep}). All other qualitative conclusions of the paper
are valid in a wider class of semimetals. The obtained collective
modes can be observed using the by now standard plasmonics
experiments, see
Refs.~\onlinecite{bas,kop17,kop18,fei12,chen12,polkop20,pol20pl,kop20p1,kop20p2}.

\section*{Acknowledgments}

The authors are grateful to U. Briskot, A.D. Mirlin, J. Schmalian,
M. Sch\"utt, and A. Shnirman for fruitful discussions. This work was
supported by the German Research Foundation DFG within FLAG-ERA Joint
Transnational Call (Project GRANSPORT), by the European Commission
under the EU Horizon 2020 MSCA-RISE-2019 program (Project 873028
HYDROTRONICS), and by the Russian Science Foundation, Grant
No. 17-12-01182 c (IG). BNN acknowledges the support by the MEPhI
Academic Excellence Project, Contract No. 02.a03.21.0005.

\bibliography{viscosity_refs}

\end{document}